\documentclass[useAMS,usenatbib]{mn2e}

\usepackage{psfig, epsf, epsfig}

\title[Formation of N-rich stars]{Formation of N-rich field stars in 
the high-density building blocks of the Galactic bulge}
\author[K. Bekki]
{Kenji Bekki${}^1$\thanks{E-mail:
kenji.bekki@uwa.edu.au} \\
${}^1$ICRAR M468
The University of Western Australia
35 Stirling Hwy, Crawley
Western Australia 6009, Australia}

\begin{document}

\date{Accepted, Received 2005 February 20; in original form }

\pagerange{\pageref{firstpage}--\pageref{lastpage}} \pubyear{2005}

\maketitle

\label{firstpage}

\begin{abstract}

Recent observational studies of the Galactic bulge by APOGEE
have revealed that
about 1\% of the bulge stars have 
rather high nitrogen abundances ([N/Fe]$>0.5$).
We here numerically investigate in what physical conditions
these N-rich stars (NRS)
can be formed in  spherical and disky stellar systems 
with stellar masses of $10^7-10^9 {\rm M}_{\odot}$ that are the bulge's
building
blocks. 
The principal results are as follows.
A large fraction ($>0.5$) 
of new stars formed from interstellar medium polluted (ISM) by
ejecta of  asymptotic giant branch stars
can have [N/Fe]$>0.5$ within stellar systems,
if the gas mass fraction of ISM ($f_{\rm g}$) is low ($\le 0.03$).
The mass fraction of NRS  among all stars ($f_{\rm nrs}$)
can be higher than $\approx 1$\% within $\approx$ 0.5 Gyr timescale
of star formation, if the mean
stellar densities ($\rho_{\rm s}$) of
the systems are higher than  $\approx 0.1 {\rm M}_{\odot}$ pc$^{-3}$.  
The [N/Fe] distributions   depend on
$\rho_{\rm s}$,  $f_{\rm g}$, and age distributions of their host stellar systems.
NRS have compact and disky spatial distributions within their host systems and
have rotational kinematics.
Based on these results, we propose that the 
vast majority of the bulge's NRS originate not from
globular clusters (GCs) but from 
its high-density
building blocks.
We suggest that NRS in the Galactic stellar halo have 
the same origin as those in the bulge.
We also suggest that low-density dwarf spheroidal and gas-rich dwarfs
are unlikely to  form NRS.
GCs are not only the formation sites of NRS.
\end{abstract}


\section{Introduction}

Recent observational studies of chemical abundances of the Galactic bulge
stars by APOGEE have discovered field stars with rather high nitrogen
abundances with [N/Fe]$>0.5$ (Schiavon et al. 2017, S17).
Since the observed  C, N, and Al abundances of
these N-rich stars (referred to as ``NRS'' from now on)
are very similar to those
of the so-called second generation (2G) stars 
in globular clusters (GCs) of the Galaxy
(e.g., Carretta et al. 2009),
the authors suggested that the origin of NRS is closely related to the
destruction (disintegration) of globular clusters (GCs) within the Galactic bulge
(S17). 
However, this GC destruction scenario for NRS can explain the observed
significant fraction of NRS among the bulge stars ($f_{\rm nrs} \sim 1$\% for
the bulge mass of $10^{10} {\rm M}_{\odot}$; S17), if $\approx 1000$ GCs with
a typical total mass of 2G stars being $10^5 {\rm M}_{\odot}$
per GC
(e.g., Carretta et al. 2009)
are completely destroyed to release their inner 2G stars to
the bulge region.

The required large number of destroyed GCs appears to be highly
unrealistic, given that the present-day number of GCs ($\approx 150$;
Harris 1996). This can be dubbed as ``over destruction problem'' in this
scenario.
S17 first pointed out this over destruction problem and
suggested that these NRS originated from the earlier generations of 
stars formed in the central region of the Galaxy.
The observed apparently unimodal [Fe/H] distribution
with the peak location around [Fe/H]$\sim -1$ was also suggested
to be hardly explained by simple GC destruction processes,
because the Galactic GC system is known to have a bimodal 
[Fe/H] distribution (S17).
Thus, the observed properties of NRS strongly suggest that
GCs are not the major host stellar systems of NRS in the bulge.

Previous one-zone chemical evolution models of the Galaxy showed
that [N/Fe] can become maximum ($\approx 0.2$)
around [Fe/H]$\approx -0.2$ 
(e.g., Fig. 9 in Chiappini et al. 2003). Therefore, 
the observed 
[N/Fe] of NRS appears to be hard to be reproduced in previous models in which
N-rich ejecta from stars are mixed well with interstellar medium (ISM). 
There are two possible mechanisms for the formation of NRS: one is
star formation almost ``directly'' from N-rich ejecta
of  asymptotic giant branch (AGB)  stars (``AGB scenario'').
Here new stars are formed  from gas from AGB stars without
mixing with ISM, after 
the gas is cooled down and trapped gravitationally by stellar systems.
The other is
star formation from gas polluted heavily by N-rich stellar winds of massive
OB stars (``OB wind scenario'').
Such NRS can be formed from ejecta from AGB stars, if the ejecta is not 
so much diluted by N-poor interstellar medium (ISM) during the formation of GCs
(e.g., Fig. 1 in Bekki et al. 2007, B07). Furthermore, such NRS can show
anti-correlations between N and C and between Al and Mg, depending on the models
of AGB yields (B07). 
Our previous hydrodynamical simulations of star formation
in GC-forming giant molecular clouds 
 demonstrated that (i) NRS can be formed
from gas heavily polluted by stellar winds of massive OB stars before
the explosion of Type II supernovae (SNe II) and (ii) these stars
show higher $Y$ and CN anti-correlation (Bekki \& Chiba 2007; BC07).
Thus, these two promising
mechanisms would need to be investigated in the context
of the Galactic bulge formation.

It should be noted here that NRS were discovered not only in the Galactic
bulge but also in the stellar halo (e.g., Martell \& Grebel 2010, MG10;
Koch et al. 2019).
As discussed by these authors, the origin of NRS in the halo can be understand
in the context of GC destruction like the NRS of the bulge.
However, the required number of GCs to reproduce the observed mass
fraction of NRS (CN-strong stars) in the halo ($\approx 200$)
is again a bit too large
(e.g., MG10).
Therefore, it is fair to say that  the origin of NRS in the halo is yet to
be fully understood.
If NRS both in the Galactic halo and bulge were all formed from GC
destruction, then the Galaxy should have almost 550 GCs ($\approx 400$ destroyed
and $\approx 150$ survived) initially, which appears to be
quite unreasonable (``over-population problem''; Koch et al. 2019).

\begin{table}
\centering
\begin{minipage}{80mm}
\caption{
Description of physical meanings for symbols often used 
for galactic building blocks in the present study. New stars (ns)
include all stars formed from gas (with different [N/Fe]).
}
\begin{tabular}{ll}
{Symbol}
& {Physical meaning}\\
NRS (nrs)  & N-rich stars with [N/Fe]$>0.5$  \\
ns  & new stars formed from gas  \\
$M_{\rm s}$ &  total  mass of old stars \\
$M_{\rm ns}$ &  total  mass of new stars (``ns'') \\
$M_{\rm nrs}$ &  total  mass of NRS \\
$M_{\rm g}$ &  total gas mass   \\
$M_{\rm g,a}$ & total gas mass accumulated in a system  \\
$f_{\rm nrs}$   &  mass fraction of NRS ($M_{\rm nrs}/M_{\rm s}$) \\
$f_{\rm ns}$   &  mass fraction of new stars ($M_{\rm ns}/M_{\rm s}$) \\
$f_{\rm g}$ &  gas mass fraction    \\
$R_{\rm nrs}$   &  mass ratio of NRS to new stars ($M_{\rm nrs}/M_{\rm ns}$) \\
$R_{\rm s}$ &  size of a stellar system \\
$R_{\rm e}$ & half-mass radius  \\
$\Sigma_{\rm g}$ & surface gas density  \\
$\rho_{\rm s}$ & mean stellar density within $R_{\rm s}$ \\
$\rho_{\rm e}$ & mean stellar  density within $R_{\rm e}$  \\
$\rho_{\rm s,th}$ &  threshold $\rho_{\rm s}$ for NRS formation  \\
$\rho_{\rm g,th}$ &  threshold gas density for star formation  \\
$f_{\rm rot}$   & fraction of rotational energy \\
$r_{\rm d}$   & dilution radius (for gas mixing)  \\
$\epsilon_{\rm g}$   & gravitational softening length  \\
$t_{\rm a}$   & mean stellar age (w.r.t. the start of a simulation)  \\
$\Delta t_{\rm a}$   & duration of star formation   \\
\end{tabular}
\end{minipage}
\end{table}

%%%%% TABLE1
\begin{table}
\centering
\begin{minipage}{90mm}
\caption{A summary for parameter values in the representative models.
The first character of each model ID (``S'' or ``D'') describes
whether the system is spherical or disky. The second character
in the ID
(``A'' in SA1) indicates the adopted star formation model.
The initial total stellar mass, size, and gas mass fraction
are denoted as $M_{\rm s}$, $R_{\rm s}$, and $f_{\rm g}$,
respectively. The low surface brightness galaxy model is 
denoted as ``LSB''.}
\begin{tabular}{lllll}
Model & $M_{\rm s}$ (${\rm M}_{\odot}$) & $R_{\rm s}$ (kpc)  & $f_{\rm g}$ 
& comments \\
SA1 & $10^9$ & 1.0 & 0.003 & fiducial \\
SA2 & $10^9$ & 0.7  & 0.003 & \\
SA3 & $10^9$ & 1.3  & 0.003 & \\
SA4 & $10^9$ & 2.0  & 0.003 &  \\
SA5 & $10^9$ & 2.5  & 0.003 & \\
SA6 & $10^9$ & 3.0  & 0.003 & \\
SA7 & $10^9$ & 3.5  & 0.003 & \\
SA8 & $10^9$ & 4.0  & 0.003 & \\
SA9 & $10^9$ & 5.0  & 0.003 &  \\
SA10 & $10^9$ & 1.0 & 0.0003 & \\
SA11 & $10^9$ & 1.0 & 0.001 & \\
SA12 & $10^9$ & 1.0 & 0.01 & \\
SA13 & $10^9$ & 1.0 & 0.03 & \\
SA14 & $10^9$ & 1.0 & 0.1 & \\
SA15 & $10^9$ & 1.0 & 0.3 & \\
SA16 & $10^9$ & 1.0 & 0.003 & $r_{\rm d}=\epsilon_{\rm g}$ \\
SA17 & $10^9$ & 1.0 & 0.003 & No dilution \\
SA18 & $10^9$ & 1.0 & 0.003 & $\rho_{\rm g,th}=10^2$  atom cm$^{-3}$ \\
SA19 & $10^9$ & 1.0 & 0.003 & $\rho_{\rm g,th}=10^4$  atom cm$^{-3}$ \\
SA20 & $10^9$ & 1.0 & 0.003 & $\rho_{\rm g,th}=10^5$  atom cm$^{-3}$ \\
SA21 & $10^{10}$ & 3.2 & 0.003 &  \\
SA22 & $10^{10}$ & 10.0 & 0.003 &  \\
SA23 & $10^{10}$ & 3.2 & 0.1 &  \\
SA24 & $2 \times 10^8$ & 0.1 & 0.003 & UCD model  \\
SA25 & $2 \times 10^8$ & 1.0 & 0.003 &  \\
SA26 & $10^8$ & 0.3 & 0.003 & \\
SA27 & $10^8$ & 0.3 & 0.1 & \\
SA28 & $10^8$ & 0.7 & 0.003 & \\
SA29 & $10^8$ & 1.0 & 0.003 & \\
SA30 & $10^8$ & 3.0 & 0.003 &  \\
SA31 & $10^8$ & 1.0 & 0.003 & $\rho_{\rm g,th}=10^5$  atom cm$^{-3}$ \\
SA32 & $3 \times 10^7$ & 1.0 & 0.003 &  \\
SA33 & $3 \times 10^7$ & 0.3 & 0.003 & \\
SA34 & $10^7$ & 1.0 & 0.003 &  \\
SNS1 & $10^9$ & 1.0 & 0.003 & no star formation \\
SADM1 & $10^9$ & 1.0 & 0.003 & with dark matter \\
SADM2 & $10^{10}$ & 3.2 & 0.003 &  \\
SADM3 & $10^{7}$ & 1.0 & 0.003 &  \\
SADM4 & $10^{6}$ & 1.0 & 0.003 &  \\
SAR1 & $10^{9}$ & 1.0 & 0.003 & $f_{\rm rot}=0$  \\
SAR2 & $10^{9}$ & 1.0 & 0.003 & $f_{\rm rot}=0.7$  \\
SB1 & $10^9$ & 1.0 & 0.003 & \\
SB2 & $10^{9}$ & 3.2 & 0.003 & $\rho_{\rm g,th}=10^5$  atom cm$^{-3}$  \\
SB3 & $10^{10}$ & 3.2 & 0.003 &  \\
SC1 & $10^9$ & 1.0 & 0.003 & \\
SC2 & $10^9$ & 1.0 & 0.003 & $\rho_{\rm g,th}=10^5$  atom cm$^{-3}$   \\
DA1 & $10^{10}$ & 3.2 & 0.003 & disk model \\
DA2 & $10^{10}$ & 3.2 & 0.1 & \\
DA3 & $10^{10}$ & 10.0 & 0.003 & LSB \\
DA4 & $10^{9}$ & 3.2 & 0.003 & LSB \\
DA5 & $10^{9}$ & 1.0 & 0.003 & \\
DA6 & $10^{8}$ & 1.0 & 0.003 & \\
DB1 & $10^{10}$ & 3.2 & 0.003 &  \\
DB2 & $10^{9}$ & 1.0 & 0.003 & \\
\end{tabular}
\end{minipage}
\end{table}

NRS have been observed in other types of galaxies outside the Local Group.
For example,
one of ultra-compact dwarf galaxies (UCDs) in M60
(M60-UCD1, 
``the densest galaxy'') is  
observed to have [N/Fe]=0.61,
 (Strader et al. 2013),
which clearly demonstrates that this UCD has NRS.
Furthermore   giant elliptical galaxies are observed to have
moderately high [N/Fe] (up to $\approx 0.2$) and their [N/Fe]
are higher in more luminous elliptical galaxies
(Schiavon 2007).
Given that these [N/Fe] are estimated from
integrated spectroscopic properties of their  entire populations including
both major N-normal/N-poor stars and NRS,
the observed moderately high [N/Fe] indeed implies that
they also contain a significant fraction of NRS.
Thus,
the physical understanding of the origin of NRS can possibly lead astronomers
to understand the origin of NRS in these galaxies too.

The purpose of this paper is thus to investigate 
how NRS can be formed in the Galactic bulge using our original hydrodynamical
simulations with a new model for mixing of ejecta from AGB stars.
As discussed in \S 2,  the AGB scenario is more promising than OB wind
one in the sense that the observed  $f_{\rm nrs}$
in the Galactic bulge can be more readily
reproduced in the AGB scenario without invoking an unrealistic stellar
initial mass function (IMF).
We therefore investigate (i) $f_{\rm nrs}$ and  (ii) [N/Fe] distributions
for the simulated stars
in the context of the AGB scenario.
Since the chemodynamical evolution of the Galactic bulge is very complicated and
yet to be understood (e.g. Barbury et al. 2018),
we consider two possible scenarios for the bulge formation
in the present study.
One is ``merger scenario'' in which
 the bulge is formed  from merging of massive 
building blocks of the bulge such as stellar  clumps developed
through global disk instability  (e.g., Noguchi 1998; Elmegreen et al. 2013)
 or low-mass 
dwarf-like galaxies  formed from density fluctuation
in the early universe. The other is ``in site scenario'' in which
the bulge is formed from an inner disk of the Galaxy through bar instability.

One of key parameters in the present study is the total stellar masses
($M_{\rm s}$) 
of stellar systems hosting NRS. The mass function of the building blocks
depends on whether the bulge was formed from dwarf-like galaxies or
stellar clumps (e.g., Cote et al. 2000;  Genel et al. 2012;
Mandelker et al. 2014;
Tamburello et al. 2015; Inoue \& Yoshida 2019).
For example, the latest cosmological simulations by Inoue \& Yoshida (2019)
have shown that the stellar clumps, which can be the building blocks of the
bulge, have stellar masses ranging from $10^7 {\rm M}_{\odot}$
to $10^{10} {\rm M}_{\odot}$ in disk galaxies at $z \approx 1$.
The mass function of stellar clumps in their simulations appears to be quite
flat (see their Fig. 5), though the number of massive stellar systems
with $M_{\rm s}>10^9 {\rm M}_{\odot}$ is small.
If the mass function of dwarf-like building blocks is described as
a power-law function with the slope of $-2$ like the
Local Group dwarf galaxies (e.g., Cote et al. 2000),
then such massive stellar systems should be rare.
Considering these previous works,
we will investigate the formation of NRS for a wide range of $M_{\rm s}$.

The plan of the paper is as follows.
We describe the models for spherical and disky systems that are building
block of the bulge and the details of the new numerical simulations
in \S 2.
We present the results of numerical simulations of
the formation of NRS in stellar systems with different model parameters
in \S 3.
Based on these results,
we provide several implications of the present results
in the context of NRS observed in the Galactic halo and in other galaxies
in \S 4.
We do not discuss much about the latest  observational results on the kinematics
of NRS (e.g., Fern\'andez-Trincado et al. 2019a, b;
Savino \& Posti 2019), because we focus exclusively
on the formation processes of NRS in the present paper.
Although there are many recent
chemical and dynamical studies of the Galactic bulge formation
(e.g., Shen et al. 2010;
Bekki \& Tsujimoto 2011; Grieco et al. 2012;
Martinez-Valpuesta et al. 2013;
Ness et al. 2013;
Di Matteo et al. 2014;
Athanassoula et al. 2017;
Debattista et al. 2019;
Matteucci et al., 2019).
we do not discuss the results of  these in the present study.

%%%%% TABLE2
\begin{table}
\centering
\begin{minipage}{90mm}
\caption{A summary for  star formation  models.}
\begin{tabular}{lll}
SF model & $t_{\rm a}$ (Gyr) & $\Delta t_{\rm a}$ (Gyr)  \\
& comments \\
A & 1 & 0.5 \\
B & 0.6 & 0.5 \\
C & 1 & 0.25 \\
D & 2 & 0.5 \\
E & 2 & 1.0 \\
\end{tabular}
\end{minipage}
\end{table}

\section{The model}

\subsection{NRS formation in the Galactic building blocks}

In the present study, we adopt a scenario in which 
 the Galactic bulge can be formed
from merging of galactic building blocks that are either
stellar clumps developed in the young disk of the Galaxy
(e.g., Noguchi 1998; Elmegreen et al. 2013; Inoue \& Saitoh 2012;
Tamburello et al. 2015; Bounaud 2016) or dwarf-like galaxies.  
Many previous works  investigated the formation and evolution
of galactic bulges through minor and major mergers of galactic
building blocks (e.g., Brooks \& Christensen
2016 for a recent review).
In this scenario,
NRS can be formed in these building blocks, and during the merging
of the building blocks, they can become the  member stars of the bulge.
In the previous works of the bulge formation in clumpy galaxies,
the formation of NRS in stellar clumps and mergers
was not investigated at all.

We therefore investigate
the formation of NRS in stellar systems with the total stellar
masses of $10^7-10^9 {\rm M}_{\odot}$ (e.g.,Inoue \& Saitoh 2012
for stellar clumps)
various sizes (e.g., Tamburello et al. 2015)
by adopting somewhat idealized models for the chemodynamical evolution
of the systems.  
Recent hydrodynamical simulations of the Galaxy formation have demonstrated that
the origin of the observed bimodal distribution of the Galactic disk stars on
the $[\alpha/{\rm Fe}-{\rm [Fe/H]}$ map can be naturally explained
in the context of star formation of massive clumps in the early
dynamical history of the Galaxy (Clarke et al. 2019). Fig. 18 of their
paper shows that the clump masses in the Galaxy 
range from $3 \times 10^7 {\rm M}_{\odot}$ to $10^{10} {\rm M}_{\odot}$,
which are similar to other recent works. The ``Clump 2'' in their Fig. 17
shows a bursty star formation with the duration of $\approx 0.1$ Gyr followed
by a low level of star formation: this Clump 2 can correspond to the present
better models for the formation of NRS.
Brook et al. (2004) showed that the Galactic thick disk can be formed from
hierarchical  merging of massive stellar clumps at high $z$. The stellar clumps
in their simulations (see Fig. 4 of their paper) can also correspond to the
host stellar systems of NRS.

Self-consistent simulations that describe both
(i) the clump formation within disks 
and dwarf galaxy formation  in cosmological simulations and
(ii) the NRS formation within clumps and dwarfs
should be done in our future works, because it is currently a time-consuming
and numerally costly task to do owing the necessary new mixing method
for ISM and AGB ejecta.
For comparison, we also investigate the formation of NRS in the context
of the bulge formation from a thin stellar disk through bar instability.
In this scenario,  NRS can be formed in the central region
of the initial thin disk under some physical conditions.
For convenience,
the physical meanings for symbols that are often used for the Galactic
building blocks are summarized in Table 1.

\subsection{Basic mechanisms: AGB or OB wind scenarios}

Nitrogen abundances in ejecta from AGB and stellar winds of OB stars 
can be  rather high ([N/Fe]$>0.5$). Accordingly,
if these ejecta can be converted into new stars without much  dilution
by N-poor ISM,  the stars can have high [N/Fe]. 
Such formation processes of NRS were already discussed 
by previous works on the chemical evolution of forming GCs
in the context of the AGB scenario
(e.g., B07; D'Ercole et al. 2010) and the OB wind one (BC07).
We first try to discuss which of the two is more promising in explaining
the origin of NRS in the Galactic bulge and thus is worth a detailed
investigation.
The most crucial observational constraint on any theory
of NRS formation  is $f_{\rm nrs} \approx 0.02$
in the Galactic bulge (S17). The total mass of ejecta of AGB stars
in a stellar system can be 
as large as $\approx 10$\% of the total mass of the stellar system
(e.g., see Fig.1 in Bekki 2011; B11). Therefore, if
these N-rich  AGB ejecta
can be converted into new stars, then the observed $f_{\rm nrs}$ can be
readily reproduced in the AGB scenario.

On the other hand, BC07 showed that $f_{\rm nrs}$ is rather small
($<10^{-3}$), even if the top-heavy IMF with the power-law slope
of $-1.35$ is adopted for star formation in the OB wind scenario. 
This is mainly because although a large fraction ($\approx 0.5$) of ISM
can have  high [N/Fe] ($>0.5$) due to chemical pollution
by N-rich stellar winds of first-generation OB stars,  such N-rich gas cannot be
converted efficiently into new stars owing to the strong feedback
effects of OB stars and SNe II. Possibly, $f_{\rm nrs}$ can be
significantly higher if these feedback effects are severely suppressed
in some particular environment. However, 
it seems highly unlikely that $f_{\rm nrs}$ can be
as high as $\approx 0.01$ in this OB wind scenario. 
We therefore conclude that the AGB scenario 
is more promising than the OB wind one and thus needs to be investigated
exclusively in the present paper.

A key question in the AGB scenario is how AGB ejecta cannot be diluted
by N-poor ISM to a large extent  in galaxy environments. In normal
star-forming, gas-rich galaxies,  AGB ejecta can be mixed well
and rapidly with ISM
so that [N/Fe] of the ISM cannot increase significantly. Furthermore,
N-poor ejecta from SNe II (e.g., [N/Fe]$\approx -0.8$ for [Fe/H]$\sim -1$) can 
also reduce [N/Fe] of the ISM significantly. Therefore,
if dilution of AGB ejecta by ISM and reduction of [N/Fe] by SNe II are both
severely suppressed,  NRS can be possibly formed almost directly from 
AGB ejecta. Since the wind velocity of AGB ejecta is rather low
($\approx 10$ km s$^{-1}$), it is reasonable and realistic to assume 
that these AGB ejecta can be all retained in galaxy-scale stellar systems.

\subsection{Necessity to adopt the new AGB particle method}

In almost all particle-based
chemodynamical simulations of galaxy-scale systems like
our own ones (e.g., Bekki 2013, B13), 
ejecta from  a  star (e.g., SNe II and AGB star)
is assumed to be shared by its surrounding (neighboring)
gas particles with the  total number  being several tens.
Accordingly, the increase of the total mass of nitrogen
($\Delta M_{\rm N}$) for a  gas particle around one AGB star is 
as follows:
\begin{equation}
\Delta M_{\rm N} = \frac{ M_{\rm ej, N} }{ N_{\rm nei} },
\end{equation}
where $M_{\rm ej, N}$ and $N_{\rm nei}$ are the total mass of 
nitrogen ejected from the AGB star and the total number of gas particle
around the star, respectively.
In these simulations, one stellar particle is assumed to consist
of stars with different masses and it can eject gas as SNe I, SNe II,
and AGB stars with different masses at different times.
The mass fraction of AGB ejecta in one  star particle ($f_{\rm agb}$)
is time-evolving  and small ($<0.1$).
Therefore, if [N/Fe]=0 for all gas particles,
[N/Fe]=1 for AGB ejecta, $N_{\rm nei}=50$, and $f_{\rm agb}=0.1$,
then [N/Fe] in one neighboring gas particle
can increase [N/Fe] from 0 to 0.26.

Even in this extreme case of $f_{\rm agb}=0.1$ (i.e., all gas
from all AGB stars can be ejected simultaneously $-$ an unrealistic 
assumption), the maximum possible increase of [N/Fe] is not enough to explain
the observed [N/Fe] of NRS.
In real simulations,  $f_{\rm agb}$ is considered
for different mass ranges of AGB stars 
so that the increase of [N/Fe] can be much smaller 
than the above 0.26: this can be dubbed ``numerical dilution''
in particle-based hydrodynamical simulations. 
In one-zone chemical evolution models, this artificial dilution can
be more serious, because AGB ejecta is assumed to be mixed well with
all gas in a galaxy.
Bekki (2019, B19) has adopted a new ``AGB particle'' method in which
new AGB particles can be ejected from a star when the star enters into
its AGB phase. In this AGB particle method, a new star can be converted 
directly from an AGB particle so that the  numerical dilution
problem can be avoided.
This AGB particle method has been recently adopted by Bekki \& Tsujimoto
(2019, BT19) 
in which rather high [N/Fe] ($\approx 1$) observed in the 
Galactic GC $\omega$ Cen
can be reproduced well. 
It should be noted here that $\omega$ Cen  has long been considered
to be the nucleus of a defunct dwarf galaxy that is one of building blocks
of the Galaxy
(e.g., Freeman 1993).
Thus we here adopt the new AGB particle method in our hydrodynamical
simulations for the formation of NRS.

\subsection{Initial stellar systems}

We assume that an initial stellar system
has either a Plummer spherical density profile
(e.g., Binney \& Tremaine 1987) or an exponential
disk profile  with a total stellar mass ($M_{\rm s}$),
and a size ($R_{\rm s}$).
In a Plummer model,
the scale length ($a_{\rm s}$) of the system is determined by the formula
\begin{equation}
a_{\rm s} = GM_{\rm s}/6{{\sigma}_{\rm s}}^{2}, \;
\end{equation}
where G is the gravitational constant and
${\sigma}_{\rm s}$ is 
a central
velocity dispersion.
The above equation is appropriate for systems with no initial angular
momentum (i.e., dynamically supported only by velocity dispersion).
However, it is highly likely that the building blocks of the Galactic
bulge have global rotation, which should be considered in the present study.

We therefore introduce the parameter, $f_{\rm rot}$, which
is the ratio of rotational energy ($T_{\rm rot}$)
to total kinetic energy ($T_{\rm kin}$) in a stellar system
as follows:
\begin{equation}
f_{\rm rot}=\frac{ T_{\rm rot} }{ T_{\rm kin} }.
\end{equation}
Here we assume that all stars in a system has an angular speed of $\Omega$
and thereby estimate $\Omega$ for $T_{\rm rot}=f_{\rm rot}T_{\rm kin}$. 
Since the rotational energy of a stellar particle is simply
$0.5 \times m_{\rm s} {\Omega}^2 R^2$, where $m_{\rm s}$ 
and $R$ are the stellar mass and  the distance
of the star from the center of its host stellar system, respectively,
$\Omega$ can be easily derived for a given $f_{\rm rot}$.
Each component of the  3D velocity of a star is reduced by a factor of 
$(1-f_{\rm rot})^{0.5}$ so that the total kinetic energy
of the system can be the same as the initial value.
Although we investigate the models $f_{\rm rot}=0$, 0.3, and 0.7,
we shows almost exclusively the results of those with
$f_{\rm rot}=0.3$. This is because the present key results
($f_{\rm nrs}$) do not depend so strongly on $f_{\rm rot}$:
we focus on the importance of other parameters in the formation of NRS.
Models with different $M_{\rm s}$ and $R_{\rm s}$,
which combine to determine the mean stellar density ($\rho_{\rm s}$),
are investigated in the present study so that the importance of 
$\rho_{\rm s}$ can be clearly elucidated.

In an exponential disk model,
the radial ($R$) and vertical ($Z$) density profiles of the stellar disk are
assumed to be proportional to $\exp (-R/a_{s}) $ with scale
length $a_{s} = 0.2R_{\rm s}$  and to ${\rm sech}^2 (Z/Z_{\rm s})$ with scale
length $Z_{\rm s} = 0.04R_{\rm s}$, respectively.
In addition to the
rotational velocity caused by the gravitational field of disk,
the initial radial and azimuthal
velocity dispersions are assigned to the disc component according to
the epicyclic theory with Toomre's parameter $Q$ = 1.5
so that the disk
can be stabilized against axisymmetric gravitational instability.

The stellar disk is assumed to be embedded in a massive dark matter halo
with the density profile described by 
the so-called NFW
profile (Navarro, Frenk \& White 1996) derived  from $\Lambda$CDM simulations
The adopted NFW profile is 
as follows:
\begin{equation}
{\rho}(r)=\frac{\rho_{0}}{(r/r_{\rm s})(1+r/r_{\rm s})^2},
\end{equation}
where  $r$, $\rho_{0}$, and $r_{\rm s}$ are
the spherical radius,  the characteristic  density of a dark halo,  and the
scale
length of the halo, respectively.
The $c$-parameter ($c=r_{\rm vir}/r_{\rm s}$, where $r_{\rm vir}$ is the virial
radius of a dark matter halo) and $r_{\rm vir}$ are chosen appropriately
as 16 for the low-mass dwarf with the total mass of $10^{10} {\rm M}_{\odot}$
and $r_{\rm s}=1.2$ kpc. A number of spherical systems are also assumed
to have dark matter.

A  stellar system is formed in a single starburst with
the canonical Salplter IMF with the power-law slope ($\alpha$) of $-2.35$,
the lower mass cut-off ($m_{\rm l}$) of $0.1 {\rm M}_{\odot}$,
and the upper mass cut-off ($m_{\rm u}$) of $50 {\rm M}_{\odot}$.
The stellar system is assumed to have a mean age of $t_{\rm a}$ and 
an age dispersion of $\Delta t_{\rm a}$: this $t_{\rm a}$ is 
an age with respect to the starting time of a simulation (not
an age with respect to the present time). 
This $\Delta t_{\rm a}$ corresponds to the duration of the starburst in the
stellar system.
The age distribution of stars in the system
is assumed to be  a Gaussian with
a mean age of $t_{\rm a}$ and 
a dispersion of $\Delta t_{\rm a}$.
We mainly investigate the models with $t_{\rm a}=1$ Gyr 
and $\Delta t_{\rm a}=0.5$ Gyr, though other models with
different combinations of $t_{\rm a}$ and $\Delta t_{\rm a}$ are
also investigated.

\subsection{Gas and star formation}

We investigate star  formation from AGB ejecta mixed with ISM
using our original chemodynamical simulations codes that can be run
on GPU clusters (B13, Bekki 2015). 
The code
combines the method of smoothed particle
hydrodynamics (SPH) with calculations of three-dimensional
self-gravitating fluids in astrophysics and incorporate various
physics of ISM such as metallicity-dependent radiative cooling,
chemical enrichment by SNe Ia, SNe II, and AGB stars,
dust growth and destruction, and ${\rm H_2}$ formation on dust grains.
Since the details of the code are given in B13, we just briefly describe
it in this paper.

An initial stellar system is assumed to have cold gas 
with a mass of $M_{\rm g}$ that is left
from the formation of the system itself.
The gas mass fraction ($f_{\rm g} = M_{\rm g}/M_{\rm s}$)
is a key parameter that can control the formation processes of NRS
in the system, because dilution of AGB ejecta by 
a large amount of ISM  can prevent the formation of NRS.
The initial distribution of the gas is assumed to follow the distribution
of the stellar system.
The initial [Fe/H] and [N/Fe] are  set to be $-1$
and $-0.1$, respectively, and this initial [Fe/H] corresponds to 
the peak [Fe/H] in the observed [N/Fe] distribution
of NRS in the Galactic bulge (S17).
When the mass density of a SPH gas particle
exceeds a threshold value for
star formation ($\rho_{\rm g, th}$),
then the particle is
converted into a  collisionless
``new star'' particle
as follows:
\begin{equation}
\rho_{\rm g} > \rho_{\rm g, th},
\end{equation}
where $\rho_{\rm g,th}$ is set to be  $10^{3}$ atom cm$^{-3}$
for most models.
Since the present results can depend on $\rho_{\rm g, th}$,
we will also investigate the models with
$\rho_{\rm g, th}=10^{5}$ atom cm$^{-3}$
that corresponds to
the typical mass density of the core of a GMC
(e.g.,  Bergin \& Tafalla 2007)
and those with 
$\rho_{\rm g, th}=10^{3}$ and $10^4$  atom cm$^{-3}$.

In the present study, SN feedback effects are not included 
(i.e., ``switched off'') 
for the following two reasons.
First, we focus on the global parameters of stellar systems 
(e.g., $M_{\rm s}$, $\rho_{\rm s}$) that can control the [N/Fe]
distributions of new stars rather than on details of SN feedback
effects in this preliminary study.
Second, the star formation rates (SFRs) in most models are rather
low ($\approx [10^{-3}-10^{-2}] {\rm M}_{\odot}$ yr$^{-1}$), which implies
that the maximum possible mass of stars in the star formation
($m_{\rm u})$ is rather low (e.g., Weidner et al. 2013, W13). This means
 a rather small number of massive stars with $m>8 {\rm M}_{\odot}$
and thus a much less effect of SN feedback on ISM.
Furthermore, collisions between molecular clouds,
which can be a major trigger for the formation  of
massive stars with $m>[20-30] {\rm M}_{\odot}$ 
in gas-rich environments (e.g., Fukui et al. 2017),
are highly unlikely in gas-poor environments in dense stellar systems.
Since we recognize the important of SNe II in the evolution of [N/Fe],
we plan to investigate this issue in our future papers.

\begin{figure*}
\psfig{file=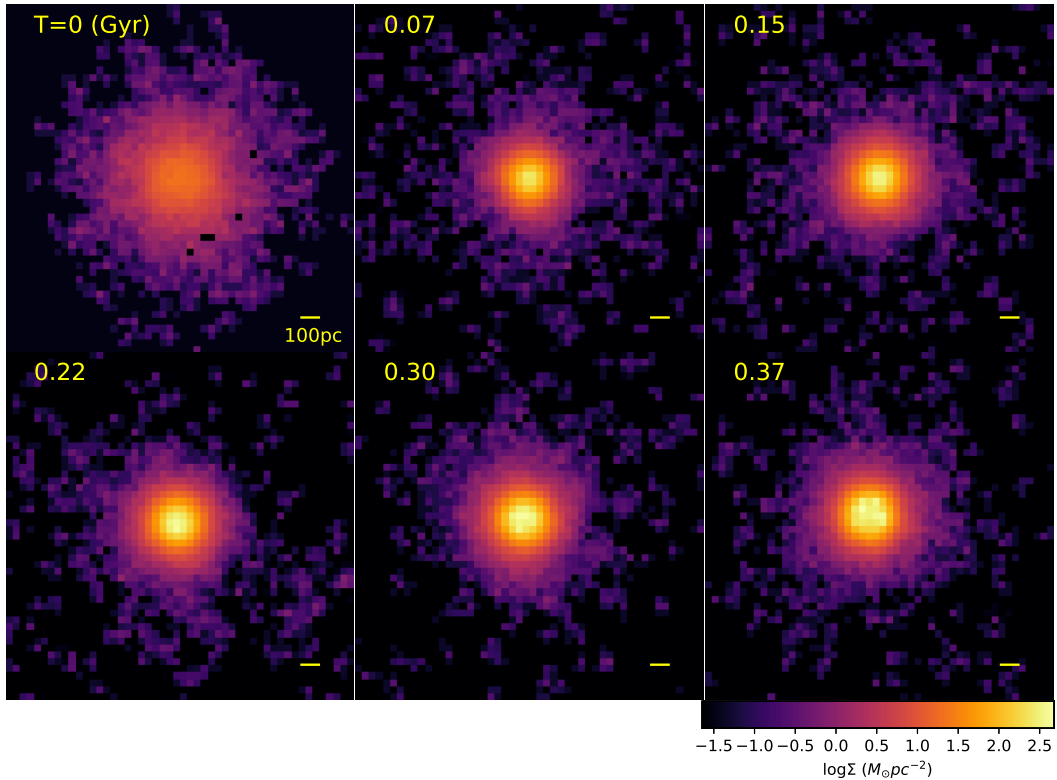,width=18.0cm}
\caption{
Time evolution of the surface mass density ($\Sigma$ in logarithmic scale)  of gas
projected onto the $x$-$y$ plane for
the model SANS1 without star formation.
Time ($T$) that has elapsed since the start of this simulation
is shown in the upper left corner of each frame.
The scale bar of 100pc pc is shown in the lower right at each panel.
}
\label{Figure. 1}
\end{figure*}

\begin{figure*}
\psfig{file=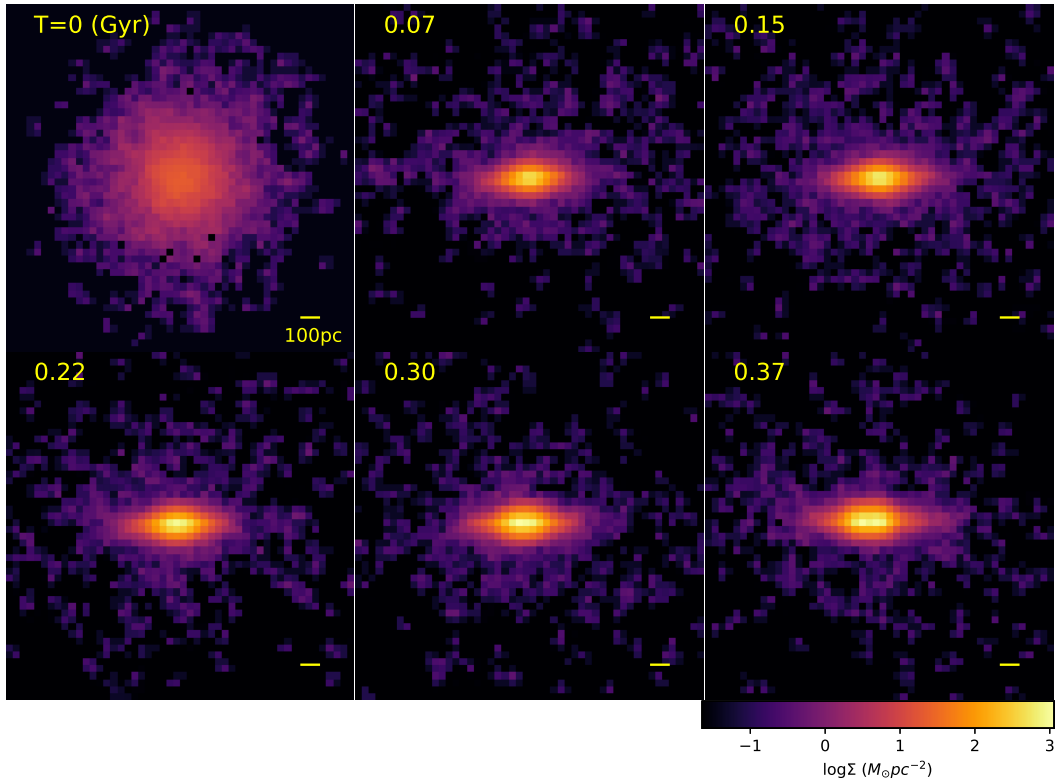,width=18.0cm}
\caption{
The same as Fig. 1 but for the $x$-$z$ projection.
}
\label{Figure. 2}
\end{figure*}

\subsection{AGB particle method}

We consider that all  AGB stars with 
$0.8 {\rm M}_{\odot} < m < 8 {\rm M}_{\odot}$
 can  eject gas with a wind velocity of
$\sim 10$ km s$^{-1}$ and the chemical abundances consistent with
the adopted chemical yield table. In the present study, we use
[N/Fe] calculated for metal-poor AGB stars by Fishlock et al. (2014, F14)
in order to estimate [N/Fe] of new stars.
In the AGB particle method (B19) adopted in the present chemodynamical simulation,
soon after a new star enters into its AGB phase,
just one new SPH gas particle (``AGB gas particle'')
is ejected from the star with its initial speed 
of 10 km s$^{-1}$ with respect to the star.
Accordingly, the present simulation
with this new mixing method is quite different from other chemodynamical simulations
of galaxies in which neighboring gas particles
around the AGB star cannot have the same chemical abundances as the AGB ejecta
because mixing of AGB ejecta and the gas particles
at the time of gas ejection  is always assumed.

In the new mixing method,
the AGB gas particle can initially have the original chemical
abundances of AGB ejecta (mixing with ISM at the time of
gas ejection is not assumed).
This N-rich AGB particle can be mixed with its surrounding ISM
and thereby change its chemical abundances.
only if there are gas particles within the ``dilution'' (``mixing'')
radius ($r_{\rm d}$, which is later described) for the AGB particle.
The search of neighboring gas particles around AGB stars is therefore
done  at each time step in a simulation.
If the gas
particle has no neighboring gas particles and have $\rho_{\rm g}$
higher than $\rho_{\rm g, th}$,  the new star formed from the gas
can have chemical abundances that are the same as those of 
AGB winds. Therefore,  new  stars formed directly
from AGB ejecta can be identified as NRS. New stars that are formed
from AGB ejecta without much dilution by ISM can also be identified
as NRS.

Each stellar particle represents a collection of stars with a particle mass:
stellar particles are not literally stars  in galaxy-scale simulations.
According, stellar particles that represent progenitor of AGB stars
can eject gas particles when they enter into their AGB phases.
The main-sequence lifetime ($t_{\rm ms}$)  is allocated for each stellar particle,
and accordingly
$i$th stellar particle  has $t_{\rm ms, \it i}$ dependent on stellar masses.
In the present study,
$T=0$ corresponds to the starting time of a simulation,
and the age of a stellar particle is measured from the starting time
:$i$th stellar particle with its age of $t_{\rm a, \it i}$ is born
at $T=-t_{\rm a, \it i}$.
Therefore, gas particles can be ejected from $i$th stellar particle when
the following condition is met:
\begin{equation}
t_{\rm ms, \it i} > t_{\rm a, \it i} + T.
\end{equation}
Therefore,  $i$th stellar particle has already died out (i.e.,  gas 
ejection has been completed) at the starting time of
a simulation, if $t_{\rm ms, \it i}$ is shorter
than $t_{\rm a, \it i}$. This can happen for stellar particles that
represent massive AGB stars with short lifetimes in some models.

We consider that AGB gas particle can change its chemical abundances
owing to diffusion of metals, if it has neighboring gas particles.
We model this diffusion process based on
the recent results from hydrodynamical simulations
of turbulent diffusion (Greif et al. 2009).
We introduce a ``diffusion radius'' ($r_{\rm d}$) within which
the chemical abundances of all particles within
$r_{\rm d}$ can be the same owing to the efficient diffusion
of metals. We mainly investigate the models with 
$r_{\rm d}=0.1 \epsilon_{\rm g}$,
where $\epsilon_{\rm g}$ is the gravitational softening length
for gas particles, which can control the spatial resolution of
a simulation.
However we investigate the models with
$r_{\rm d}=0$ (no diffusion),
$1 \epsilon_{\rm g}$ ,
and $10 \epsilon_{\rm g}$ and show only some of the results,
because $r_{\rm d}$ is not so important in the formation of NRS
in comparison with other model parameters.

\subsection{Parameter study}

We describe the results of 53 representative models with different 
parameters in the present study, and the model parameters
are summarized in Table 2. The first character in a  model 
identification (ID) number
(e.g., ``S'' in SA1) describes whether the model is a spherical (``S'')
or disky (``D'') system. The second character (e.g., ``A'' in SA1) 
in the model ID represents the ID number of the star formation model.
The parameters for  star formation models are summarized in Table 3.
For comparison,  spherical models with dark matter
are investigated, and the ID names of the models contain
``DM'' in
the third and fourth characters (e.g., SADM1).
The models without star formation (labeled as ``SNS1'') is investigated
so that we can better understand how AGB ejecta can be accreted within
stellar systems. 

Although we investigate the models with 
different $f_{\rm rot}$ (=0, 0.3, and 0.7),
we describe the results of the models with $f_{\rm rot}=0.3$ almost
exclusively,
because the present results do not depend strongly on $f_{\rm rot}$.
The models with $f_{\rm rot}=0$ (without rotation)
and 0.7 (more rotation) are labeled as SAR1 and SAR2, respectively:
all other spherical models have $f_{\rm rot}=0.3$.
We choose this $f_{\rm rot}=0.3$, because the Galactic bulge
is observed to have such rotational kinematics (e.g., Fig. 11 in Ness et al.
2013). 
The models with different $M_{\rm s}$,
$\rho_{\rm s}$ ($R_{\rm s}$), $t_{\rm a}$, $\Delta t_{\rm a}$,
$\rho_{\rm th}$, $f_{\rm g}$,
and $r_{\rm dis}$ are investigated, though  $\rho_{\rm s}$
is most extensively investigated in comparison with other parameters.

We mainly show the results of the fiducial model SA1 with 
$M_{\rm s}=10^9 {\rm M}_{\odot}$,
$R_{\rm s}=1$ kpc,
$f_{\rm g}=0.003$,
$t_{\rm a}=1$ Gyr,
$\Delta t_{\rm a}=0.5$ Gyr,
and
$\rho_{\rm g,th}=10^3$ atom cm$^{-3}$,
because this gas-poor  models shows a typical behavior of NRS formation in the present
study. The initial total numbers of particles in  a stellar system
are 
are $10^6 (1+f_{\rm g})$, where $f_{\rm g}$ is the initial gas mass fraction
in the system: $N=1100000$ for $f_{\rm g}=0.1$, for example.
The  total number of gas particles
increase dramatically with time due to the ejection of gas particles from AGB stars.
The mass and size resolutions are $10^3 {\rm M}_{\odot}$ and 5 pc,
respectively, in the fiducial model.

The disk models with 
$M_{\rm s}=10^{10} {\rm M}_{\odot}$,
$R_{\rm s}=3.2$ kpc,
$f_{\rm g}=0.003$ kpc,
$t_{\rm a}=1$ Gyr,
$\Delta t_{\rm a}=0.5$ Gyr,
corresponds to the progenitor disk of the Galactic bulge.
Although it is unlikely that this massive stellar disk is formed
from a single giant starburst, we investigate this model
in order to discuss ``in situ'' formation of NRS within the bulge's
progenitor disk.
For comparison, we also investigate a massive spherical stellar system
with $M_{\rm s}=10^{10} {\rm M}_{\odot}$,
Therefore, if the formation of NRS in these 
spherical and disky models are demonstrated in the present study,
then it means that ``in situ'' formation
of NRS is possible.

One of key properties of NRS is the mass fraction of NRS among all stars 
initially in a system, which is defined as follows:
\begin{equation}
f_{\rm nrs} = \frac{ M_{\rm nrs} } { M_{\rm s} },
\end{equation}
where $M_{\rm nrs}$ is the total mass of NRS with [N/Fe]$>0.5$.
Here we assume that there are no NRS in the old stars of a system.
In addition to $f_{\rm nrs}$,  the mass ratio of NRS to all new
stars ($R_{\rm nrs}$)  is also investigated in detail and is defined as follows:
\begin{equation}
R_{\rm nrs} = \frac{ M_{\rm nrs} } { M_{\rm ns} },
\end{equation}
where $M_{\rm ns}$ is the total mass of news stars.
If this $R_{\rm nrs}$ is quite low in a stellar system, 
it would be difficult for observations
to find NRS among a limited number of stars that can be observed for
the system.
We also investigate the mass fraction of new stars in a stellar system,
which is defined as follows ($M_{\rm ns} << M_{\rm s}$):
\begin{equation}
f_{\rm ns} = \frac{ M_{\rm ns} } { M_{\rm s} },
\end{equation}

Although we investigate stellar systems  with star formation models A$-$E,
we describe the results of the models A, B, and C in detail: those for
D and E are not described.
This is because stellar systems with the models D and E
show rather small $f_{\rm nrs}$ $<0.005$ (i.e., less than the required
0.01). In these D and E models with $t_{\rm a}=2$ Gyr,  
massive and intermediate-mass AGB stars ($m > 5 {\rm M}_{\odot}$) are
assume to have finished gas ejection at the starting time of a simulation.
Therefore, the total mass of AGB eject that can be accumulated
within a system is rather small ($<0.005$) so that $f_{\rm nrs}$ cannot
exceed the required $f_{\rm nrs}=0.01$.
This means that stellar systems should retain AGB ejecta from a wide mass
range to form a large fraction NRS$-$ a constraint for the formation of NRS
in the AGB scenario.

\section{Results}
\subsection{Massive spheroidal systems}

Figs. 1 and 2 show the 2D maps of the projected total  gas density
of ISM ($\Sigma_{\rm g}$)
at six selected time steps in the the spherical model
SNS1 without star formation
in which the parameter values  are exactly the same
as those in  the fiducial model SA1. 
Since star formation is not included in this model,
it can more clearly show how the spatial distribution
of original ISM mixed with  AGB ejecta evolves with time in
the deep potential well of the system.
Clearly, $\Sigma_{\rm g}$ gradually increases due to the 
continuous ejection of gas  from AGB stars with different masses within
the system. The system can develop
a thick disk-like structure with 
$\log \Sigma_{\rm g} \approx 2 {\rm M}_{\odot}$ pc$^{-2}$
within $\approx 0.1$ Gyr. Gas ejected from AGB stars can fall onto the
thick  gas disk so that
the disk can grow slowly.

The formation of the thick  gas disk is closely related to the initial 
angular momentum of the stellar system with $f_{\rm rot}=0.3$
(i.e., rotational energy of stars being  30\% of the total kinetic
energy of the system). Therefore, it is possible that the gas disk
become more compact for the models with smaller $f_{\rm rot}$.
As described later, new stars can be formed steadily in the gas disk
composed of original ISM and AGB ejecta so that the stars can have
disky kinematics. 

\begin{figure}
\psfig{file=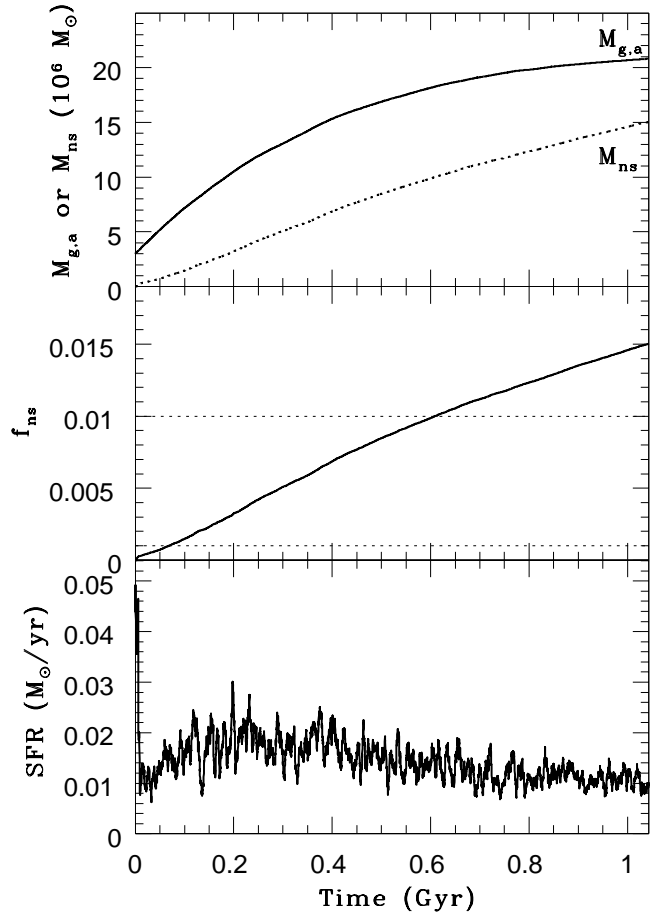,width=8.5cm}
\caption{
Time evolution of the total masses of gas and new stars (top),
the mass fraction of new stars ($f_{\rm ns}$, middle),
and the star formation rate in the spherical stellar system for the fiducial
model SA1. The total masses  of gas that is accumulated in the system
($M_{\rm g, a}$)
and new stars ($M_{\rm ns}$)  are shown by solid and dotted lines, respectively.
The upper and lower dotted lines in the middle frame
indicate $f_{\rm ns}=0.01$ and 0.001, respectively, for comparison.
}
\label{Figure. 3}
\end{figure}

\begin{figure}
\psfig{file=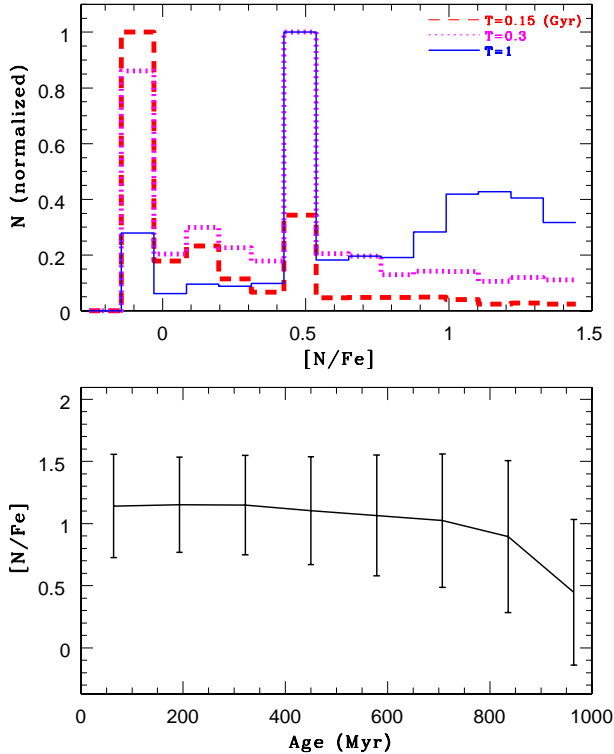,width=8.0cm}
\caption{
Simulated [N/Fe] distributions
at $T=0.15$ Gyr (red solid),
0.3 Gyr (magenta dashed), and 1 Gyr (blue solid),
in the upper panel and age-[N/Fe] relation 
at the final
time step ($T=1$ Gyr) in the lower panel
for the fiducial model SA1.
The [N/Fe] distribution is normalized by the 
number of stars in the bin that shows the maximum number of stars
just for clarity.
The error bars at each age bin indicates the $1\sigma$ dispersion of
[N/Fe].
}
\label{Figure. 4}
\end{figure}

The time evolution of gaseous distribution in the fiducial model SA1
with star formation
is essentially the same as that shown in Figs. 1 and 2 for SNS1 without
star formation
Fig. 3 shows the time evolution of (i) the total mass of gas that 
is the sum of the  original ISM 
and the total mass of gas ejected from AGB stars ($M_{\rm g,a}$),
(ii) that of new stars $M_{\rm ns}$, 
(iii)  mass ratio of new stars to the original stellar mass ($f_{\rm ns}$),
and (iv) star formation rate (SFR) in the fiducial model SA1.
The initial (spiky)  higher SFR ($ \approx 0.04 {\rm M}_{\odot}$) is due to 
star formation of original ISM that is not mixed with AGB ejecta.
This star formation corresponds to the formation of N-poor ([N/Fe]$<0$) stars.
The increase of gas mass  due to continuous ejection
of gas from AGB stars is gradual in the stellar system, 
and consequently, the SFR can be kept
low ($\approx 0.01 {\rm M}_{\odot}$ yr$^{-1}$). 
The system can have $f_{\rm ns} > 0.01$ only after $\approx 0.6$ Gyr
long continuous conversion of ISM by star formation.
The timescale for a system to have $f_{\rm nrs}$ can be 
shorter than $0.5$ Gyr in some models (e.g., 0.37 Gyr for  SB1).

\begin{figure*}
\psfig{file=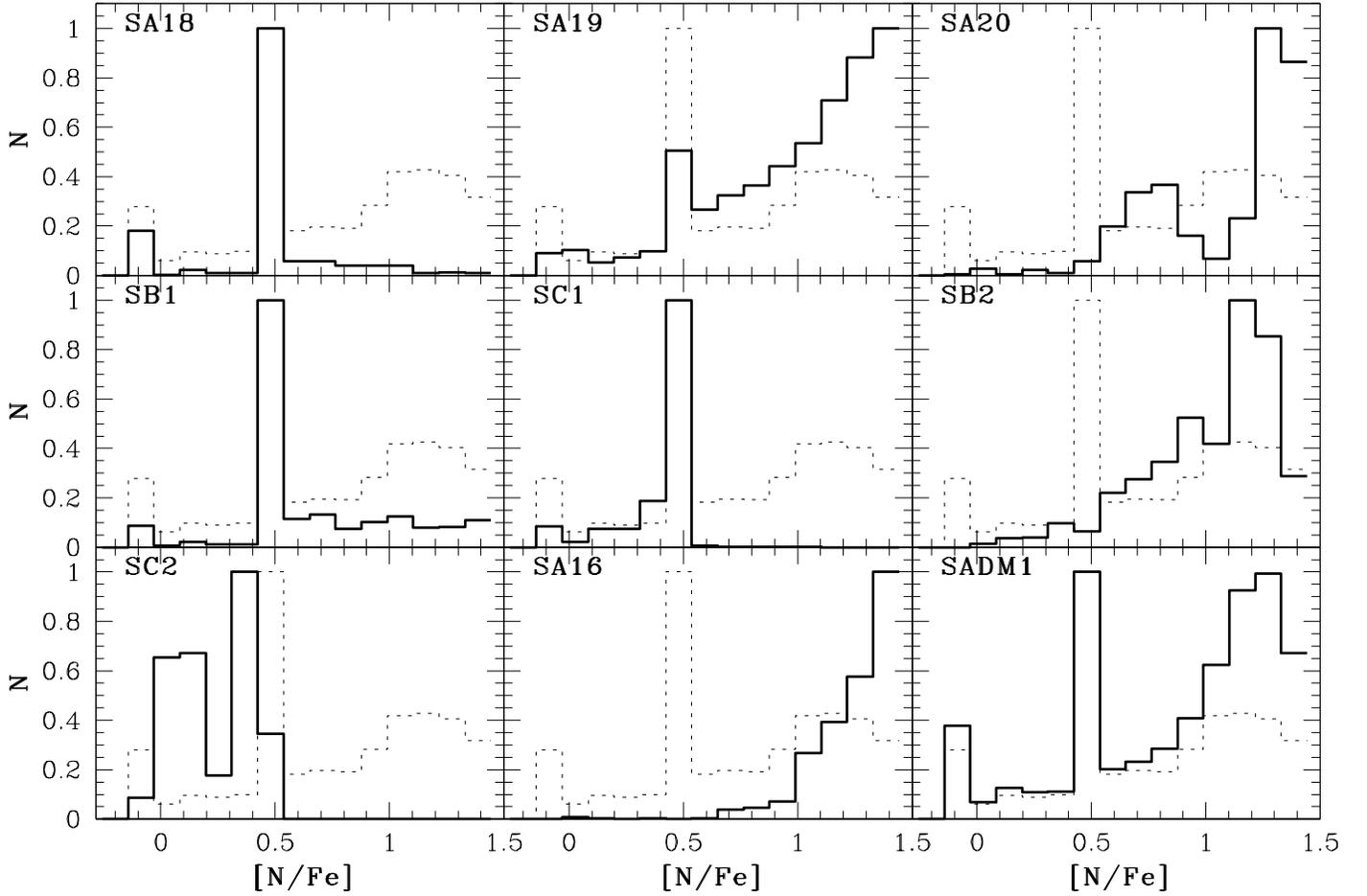,width=18.0cm}
\caption{
[N/Fe] distributions in spherical stellar models with
$M_{\rm s}=10^9 {\rm M}_{\odot}$ and  $R_{\rm s}=1$ kpc 
in which  $\rho_{\rm g, th}$, $t_{\rm a}$, and dark matter mass
are different.
Model ID is shown in the upper left corner of each panel.
For comparison, the result of the fiducial model is shown by
a dotted line in each panel.
SA18, 19, and 20 are models with the same star formation
history model ($t_{\rm a}=1$ Gyr and $\Delta_{\rm a}=0.5$ Gyr) 
as SA1 (fiducial model)
yet different $\rho_{\rm g, th}$
($10^2$, $10^4$, and $10^5$ atom cm$^{-3}$, respectively).
SB1, SB2, SC1, and SC2 are those with different star formation
histories of the stellar systems: 
$t_{\rm a}=0.6$ Gyr, $\Delta_{\rm a}=0.5$ Gyr, and
$\rho_{\rm g, th}=10^3$ atom cm$^{-3}$
for SB1,
$t_{\rm a}=0.6$ Gyr, $\Delta_{\rm a}=0.5$ Gyr, and
$\rho_{\rm g, th}=10^5$ atom cm$^{-3}$
for SB2,
$t_{\rm a}=1$ Gyr, $\Delta_{\rm a}=0.25$ Gyr,
and
$\rho_{\rm g, th}=10^3$ atom cm$^{-3}$
for SC1, and
$t_{\rm a}=1$ Gyr,  $\Delta_{\rm a}=0.25$ Gyr, and
$\rho_{\rm g, th}=10^5$ atom cm$^{-3}$
for SC2.
The two models, SA16 and SADM1, are the same as the fiducial model except that
(i)  dilution  is more efficient ($r_{\rm d}=\epsilon_{\rm g}$)
in SA16  than in SA1
($r_{\rm d}=0.1\epsilon_{\rm g}$)
and (ii) the stellar system is embedded in dark matter halo for SADM1.
}
\label{Figure. 5}
\end{figure*}

Recent theoretical studies of star formation have suggested that 
there can be a strong relation between a SFR and a maximum possible
mass of stars ($m_{\rm u}$, the upper mass cut-off in the IMF)
in a star-forming galaxy (e.g., W13). The predicted
SFR$-m_{\rm u}$ relation (e.g., Fig. 1 in Bekki et al. 2017)
implies that $m_{\rm u}$ can be $\approx 25 {\rm M}_{\odot}$  for
the derived rather low SFR.
Such a possible low $m_{\rm u}$ means a smaller number of SNe II that
can expel ISM from a stellar system owing to the energetic feedback effects.
Therefore, the rather low SFR in the fiducial model
suggests that SNe II does not influence the evolution of ISM so much.
Since [N/Fe] of SNe II can be quite small ([N/Fe]$\sim -0.8$
for low-metallicity; BT19),
the possibly very small number of SNe II means that SNe II cannot lower
[N/Fe] of ISM that is enriched by AGB ejecta.

Fig. 4 shows that there is a distinct peak around [N/Fe]$\approx 0.5$
in the [N/Fe] distribution, which reflects that a large fraction of 
AGB ejecta can be converted into new stars without much dilution by
N-poor original ISM. There is also a weak peak around [N/Fe]$\approx 1.2$,
which corresponds to star formation from gas ejected from AGB stars with
lower masses. Since the adopted [N/Fe] is rather high ([N/Fe]$>1$) for
masses lower than $5 {\rm M}_{\odot}$, new stars formed almost directly
from AGB ejecta can have rather high [N/Fe] $>1$ in the present study.
The weak peak around [N/Fe]$\sim -0.1$ is due to the formation of new
stars from original ISM with low [N/Fe].
A key physical process here is that new stars are formed from AGB ejecta
that is gravitationally trapped by the stellar system and cooled down
to form high-density gaseous regions. If AGB winds are all removed
from the systems, then the [N/Fe] distribution of new stars formed 
from ISM simply reflects the original ISM abundance (i.e., low [N/Fe]).

\begin{figure*}
\psfig{file=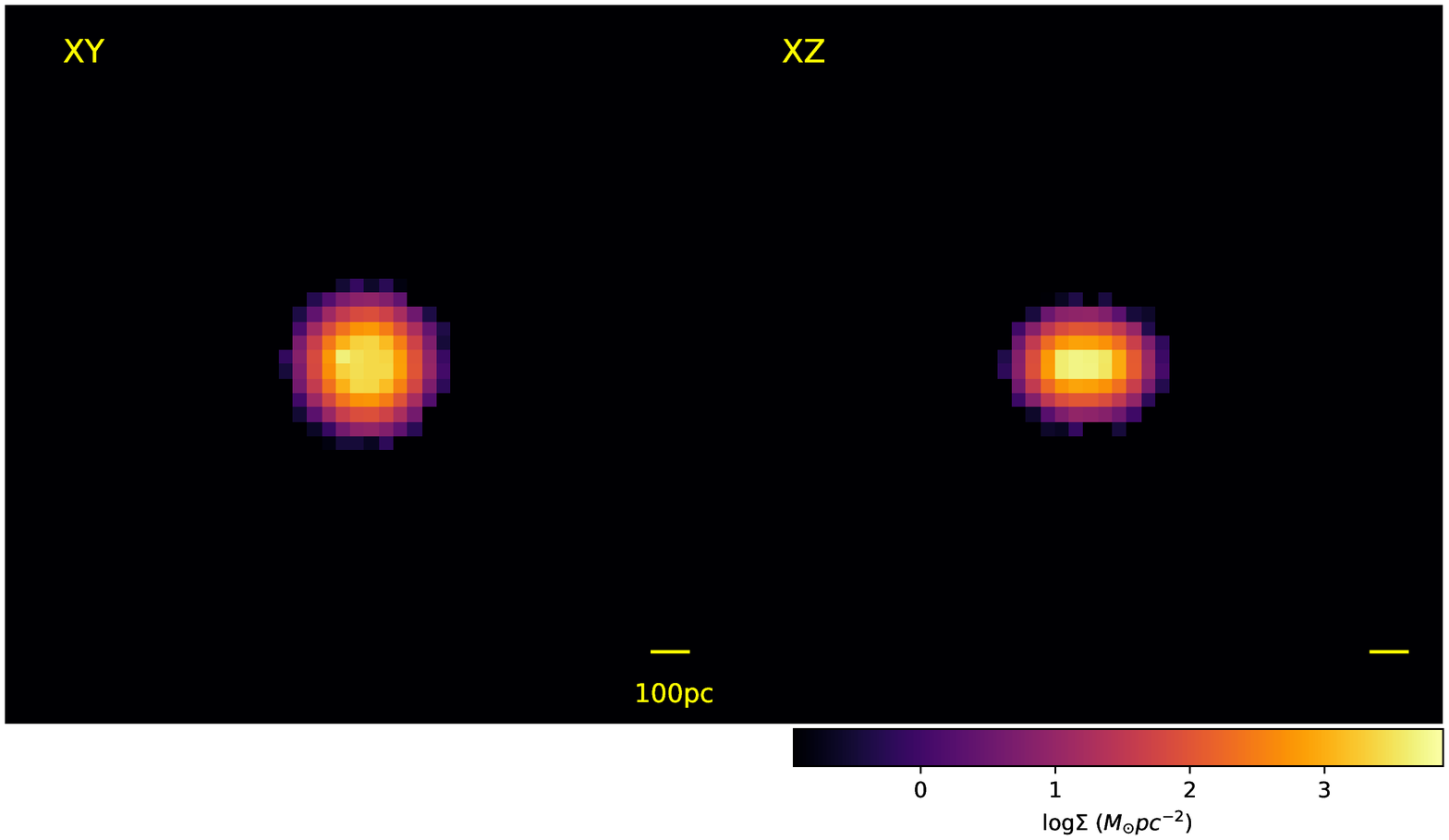,width=18.0cm}
\caption{
Final distributions of new stars projected onto the $x$-$y$ (left)
and $x$-$z$ planes at $T=1$ Gyr in the fiducial model.
Since 89\% of the stars are NRS in this model,
these spatial distributions  are very similar to those of NRS.
}
\label{Figure. 6}
\end{figure*}

The peak location of the [N/Fe] distribution reflects the mass range
of AGB stars whose ejecta is efficiently
converted into new stars without much dilution
by ISM.
The [N/Fe]$\approx 0.5$ peak
fiducial model means that ejecta from
AGB stars with $m \approx 3 {\rm M}_{\odot}$
contributes significantly to the formation of NRS.
It should be noted here, however,
that the location of the peak 
can be changed significantly, if different yield tables for
AGB stars are adopted.
For example, the predicted [N/Fe] for different masses of AGB stars
are different between F14 and Ventura et al. (2013, V13)
that has been often used in chemical evolution studies of GCs:
[N/Fe]  is $\approx 2.5$
for intermediate  AGB stars ($ m \approx 3{\rm M}_{\odot}$) 
with $Z=10^{-3}$ in V13
whereas it is $\approx 0.5$ in F14.
The significantly larger [N/Fe] for AGB winds in V13
implies that the peak position of the simulated [N/Fe] distribution
moves to the higher [N/Fe], if the present chemodynamical models
adopt the AGB yields from V13.
This further implies that the fraction of NRS ($f_{\rm nrs}$)
for lower metallicities  can be 
lower in chemodynamical simulations with AGB yields from V13,
if gas from $m=3{\rm M}_{\odot}$ AGB stars is the
major gaseous fuel for the formation
of NRS.

Since the predicted
AGB yields are significantly different over a wide range of stellar
masses between F14 and V13, it could be possible that the details of the simulated
[N/Fe] distributions can be also
different between models with AGB yields from
F14 and V13.
Currently, there is an uncertainty in the observed [N/Fe] distribution 
of the Galactic bulge owing to the upper limit of [N/Fe] in the ASPCAP
spectral library (S17).
Therefore, it is not possible whether the derived [N/Fe] peak
around $\approx 0.5$ and the shape of the [N/Fe]
distribution are consistent with the corresponding observations.
If a large number of stars with [N/Fe]$>1$ are found in future observational
studies with a better spectral library for [N/Fe]$>1$,
then they can be used for a strong constraint for theoretical models
of NRS formation. 

The age-[N/Fe] relation of new stars in this model is almost flat
for ages younger than $\approx 700$ Myr, though
the relation shows large dispersion for all 
age bins over $\approx 1$ Gyr evolution of the system.
Different local regions of  ISM are polluted by AGB winds to
different degrees so that they can have quite
different [N/Fe]
over $\approx 1$ Gyr evolution of the system: chemical pollution by AGB winds
proceeds in an inhomogeneous way.
This is a physical reason why the  dispersion of [N/Fe] in each age bin
can be quite large.
New stars 
older  than $\approx 700$ Myr
shows a decreasing [N/Fe] with increasing ages.
Mean [N/Fe] in ISM can increase slowly and steadily in the early evolution
of the system in the first few 100 Myrs. Therefore,
new stars formed from ISM can show lower [N/Fe] in the first few 100 Myrs.

Fig. 5 compares the simulated [N/Fe] distributions between nine 
representative spherical stellar systems  with
different parameters.  Structural
parameters for the stellar systems are exactly the same
as those in the fiducial model
(i.e., $M_{\rm s}=10^9 {\rm M}_{\odot}$ and $R_{\rm s}=1$ kpc),
yet other parameters (e.g., $\rho_{\rm g, th}$ are different.
The model SA18 with $\rho_{\rm g, th}=10^2$ atom cm$^{-3}$ show a much
smaller number of NRS in comparison with
SA19 with $\rho_{\rm g, th}=10^4$ atom cm$^{-3}$ 
SA20 with $\rho_{\rm g, th}=10^5$ atom cm$^{-3}$.
In the present model,  [N/Fe] of original N-poor
gas particles  can increase due to their mixing
with AGB (through dilution) before the particles are converted into
new stars. Therefore,  if the duration of ``gas phase'' is longer,
then  original N-poor gas particles can increase their [N/Fe] to a
larger extent.
Thus, the models with lower $\rho_{\rm g, th}$,  in which gas phase
is typically shorter,  can have a smaller number of NRS.

   The strongest peaks in the [N/Fe] distributions
are not around [N/Fe]$\approx 0.5$ in SA19 and SA20, 
as shown in the fiducial model,
but they are shifted to higher [N/Fe].  In these models with
rather high $\rho_{\rm th}$, there is a very small number of N-poor
stars ([N/Fe]$<0$), because original N-poor gas
particles with  initially small [N/Fe] can be converted
into new stars only after their [N/Fe] becomes rather high due to 
mixing with AGB ejecta. Although these two models SA19 and SA20
show slightly higher 
$R_{\rm nrs}$ (0.94 and 0.98, respectively) than the fiducial model SA1,
their $M_{\rm nrs}$  are
$1.1 \times 10^7 {\rm M}_{\odot}$
and $4.1 \times 10^6 {\rm M}_{\odot}$, respectively.
which are significantly lower than SA1.
Thus, the model SA20 with $f_{\rm nrs}=0.0041$, which is less
than 0.01, might not be  useful to discuss the observed $f_{\rm nrs}$ in
the Galactic bulge and halo.

The models with different $t_{\rm a}$ and $\Delta t_{\rm a}$ show 
different [N/Fe] distributions, which reflect the relative contribution
of high- and intermediate-mass AGB stars. The model with smaller
$t_{\rm a}=0.6$Gyr  ($\Delta t_{\rm a}=0.5$ Gyr) shows
a distinct peak around [N/Fe]$\approx 0.5$ and a smaller fraction
of NRS with [N/Fe]$>1$. This model also has  larger
$M_{\rm nrs}$ ($3.8 \times 10^7 {\rm M}_{\odot}$) 
and $R_{\rm nrs}$ (0.96) thus $f_{\rm nrs}= 0.038$.
This larger $f_{\rm nrs}$ is due to the participation of a wider mass
range (thus a larger mass fraction) of AGB stars in star formation:
ejecta from AGB stars with ages older than 0.1 Gyr can be mixed with ISM.
However, the model with
$t_{\rm a}=0.6$Gyr  ($\Delta t_{\rm a}=0.5$ Gyr,
and $\rho_{\rm g}=10^5$ atom cm$^{-3}$ shows the strongest peak
around [N/Fe]=1.1 and $f_{\rm nrs}=0.01$ (by a factor of $\approx 3$
lower than the above model). Again, $\rho_{\rm g, th}$ should not be
too high to construct a model that shows larger $f_{\rm nrs}$.

The models with
$t_{\rm a}=1$Gyr and $\Delta t_{\rm a}=0.25$ Gyr does not have NRS
with [N/Fe]$>1$ for two different $\rho_{\rm g, th}$. This is mainly
because star formation is assume to occur only after
N-rich ejecta of AGB stars with
masses lower than $5 {\rm M}_{\odot}$ has been removed from the stellar
systems completely.
This model would not be so appropriate for a comparison with the observed
$f_{\rm nrs}$ of the Galactic bulge,
because $f_{\rm nrs}$ is low ($0.006$ for
$\rho_{\rm g, th}=10^3$ atom cm$^{-3}$).
The model with much more efficient dilution ($r_{\rm d}=16$ pc) 
shows a very small fraction of stars with [N/Fe]$<0$,
because original N-poor gas particles are converted into new stars
only after pollution by AGB stars.
The model with dark matter halo has an almost identical [N/Fe] distribution
to the fiducial model, which suggests that the dark matter potential
does not influence the dilution process of AGB ejecta and the star
formation process in the stellar system with
$M_{\rm s}=10^9 {\rm M}_{\odot}$.

Fig. 6 demonstrates that the distribution of new stars 
at $T=1$ Gyr in the fiducial model  is rather  compact and
flattened with the central density as high as 
$10^4 {\rm M}_{\odot}$ pc$^{-2}$.
The vast majority of the new stars reside within the central 200 pc of the original stellar system. Since almost 90\% of the stars are NRS, this result
means that NRS are the dominant stellar populations in the central
region of the system. 
Fig. 7 clearly shows that NRS has rotational kinematics with
the maximum circular velocity of $\approx 100$ km s$^{-1}$ and 
a sign of cylindrical rotation in the $x$-$z$ projection, which reflect
the dissipative formation of a gas disk from accreted AGB ejecta.
This rotational kinematics can be seen in all models of the present-study,
because the original stellar systems are assumed to have initial rotation.

\begin{figure}
\psfig{file=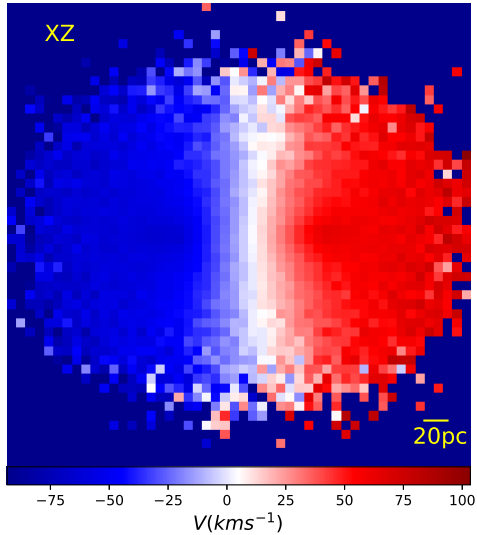,width=8.0cm}
\caption{
2D map of line-of-sight velocities of new stars projected onto the $x$-$z$ plane
at $T=1$ Gyr
in the fiducial model.
Since there are  mesh points where there is no star,
the bluest colors are allocated for such empty mesh points. 
}
\label{Figure. 7}
\end{figure}

\begin{figure}
\psfig{file=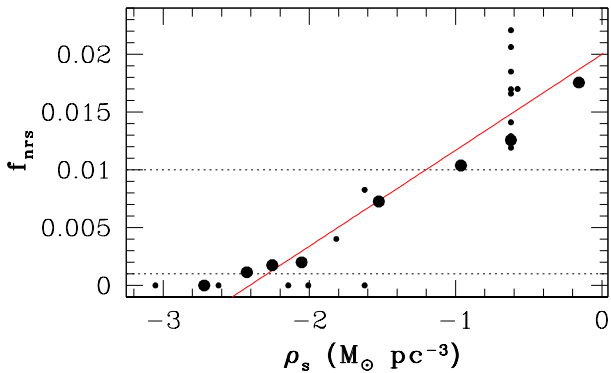,width=8.0cm}
\caption{
Mass fractions of NRS ($f_{\rm nrs}$) 
as a function of the initial mean stellar densities ($\rho_{\rm s}$)
of stellar systems for selected 25 models.
The fiducial model and those in which model parameters
other than $R_{\rm s}$ are exactly the same as those in the fiducial
model are shown as big filled circles. All other models are shown by
small filled circles.
The upper and lower dotted lines in the middle frame
indicate $f_{\rm ns}=0.01$ and 0.001, respectively, for comparison.
The derived relation between $f_{\rm nrs}$ and $\rho_{\rm s}$
($f_{\rm nrs} = 0.0083 \log \rho_{\rm s} + 0.02$)
is shown by a red solid line.
}
\label{Figure. 8}
\end{figure}

\subsection{Parameter dependence}

The present results depend on $M_{\rm s}$, $R_{\rm s}$, $f_{\rm g}$,
$t_{\rm a}$, $\Delta t_{\rm a}$, $\rho_{\rm g, th}$, and
initial stellar distributions in stellar systems.
Among these, it is found that the initial mean stellar density
of a system ($\rho_{\rm s}$) is a key parameter that controls $f_{\rm nrs}$
at the final time step of a simulation. Since observational studies of
galactic mass-size relations often derive a scaling relation between
the half-mass radius ($R_{\rm e}$) and the mean stellar density
at $R_{\rm e}$ ($\rho_{\rm e}$), we here clarify the relation between
$\rho_{\rm s}$ and $\rho_{\rm e}$ estimated from the adopted Plummer model
as follows:
\begin{equation}
\rho_{\rm e} = 32 \rho_{\rm s}.
\end{equation}
Accordingly, $\rho_{\rm e}$ is used to discuss possible $f_{\rm nrs}$
in various types of galaxies
based on the observed mass-size relations of galaxies.

We mainly describe how the present results depend on $\rho_{\rm s}$,
$f_{\rm g}$, and stellar distributions of stellar systems,
because other parameters are not so important as these three
for $f_{\rm nrs}$.
For example, the models with  (SADM1$-$2) 
and without dark matter show similar  $f_{\rm nrs}$,
and $f_{\rm nrs}$ does not depend on $f_{\rm rot}$ (SA1, SAR1, and SAR2).
The model SADM4 with $M_{\rm s}=10^6 {\rm M}_{\odot}$
shows $f_{\rm nrs}$ ($\approx 0.006$) that is quite different from 
$f_{\rm nrs}$ derived in 
the models  with $M_{\rm s}=10^6 {\rm M}_{\odot}$ 
yet without dark matter ($f_{\rm nrs}=0$).
The physical reason for this
no star formation is that AGB ejecta cannot be retained in the systems
due to their shallow gravitational potentials.
This means that low-mass dwarf galaxies need to be embedded in massive
dark matter halos to form NRS.

\begin{figure*}
\psfig{file=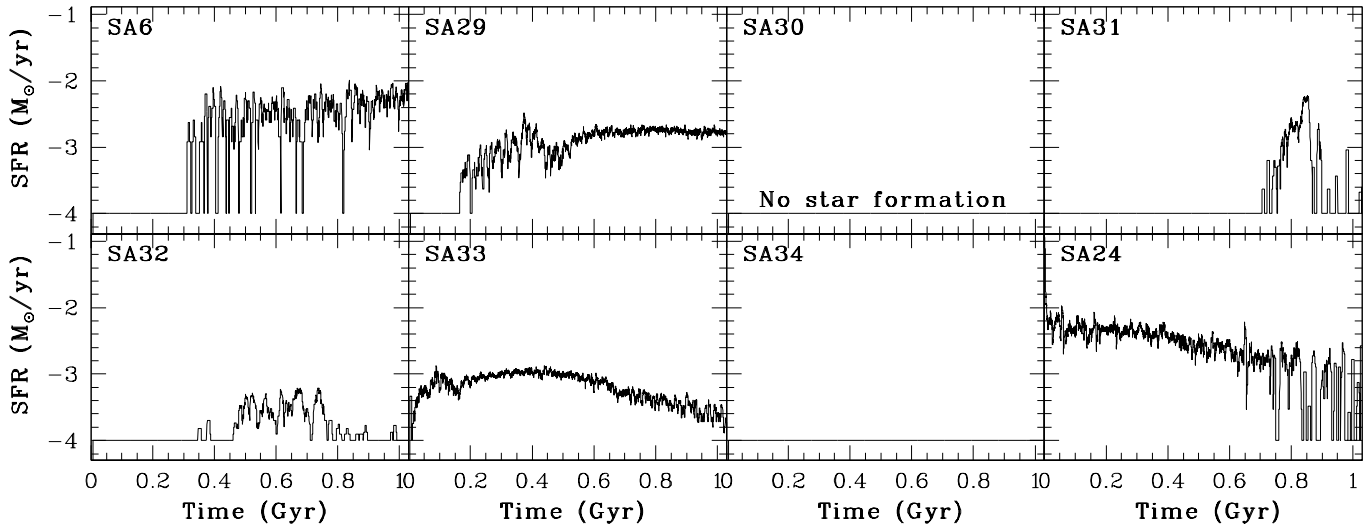,width=18.0cm}
\caption{
Evolution of star formation rates (SFRs) in the selected eight models
with different $\rho_{\rm s}$. 
Model ID is shown in the upper left corner of each panel.
For convenience, SFR=0 is plotted as $\log {\rm SFR}=-4$.
SA6, 29, 30, 31, 32, and 34 are all low-density models
with different $\rho_{\rm g, th}$:
$M_{\rm s}=10^9 {\rm M}_{\odot}$ and $R_{\rm s}=3$ kpc
for SA6,
$M_{\rm s}=10^8 {\rm M}_{\odot}$ and $R_{\rm s}=1$ kpc
for SA29,
$M_{\rm s}=10^8 {\rm M}_{\odot}$ and $R_{\rm s}=3$ kpc
for SA30,
$M_{\rm s}=10^8 {\rm M}_{\odot}$ and $R_{\rm s}=1$ kpc 
($\rho_{\rm g, th}=10^5$ atom cm$^{-3}$)
for SA31,
$M_{\rm s}=3 \times 10^7 {\rm M}_{\odot}$ and $R_{\rm s}=1$ kpc 
for SA32,
$M_{\rm s}=10^7 {\rm M}_{\odot}$ and $R_{\rm s}=1$ kpc 
for SA34.
The low-mass ($M_{\rm s}=3\times 10^7 {\rm M}_{\odot}$) and high-density
($R_{\rm s}=0.3$ kpc) model SA33 can show star formation (SF) whereas
the low-mass ($M_{\rm s}=3 \times 10^7 {\rm M}_{\odot}$) and low-density
($R_{\rm s}=1$ kpc) model SA32 can show little SF.
The low-density  ($M_{\rm s}=10^7 {\rm M}_{\odot}$ and $R_{\rm s}=1$ kpc)
model SA32,
which corresponds to dwarf spheroid,
can show no SF.
The model SA24 corresponds to an UCD (like M60-UCD)
with
$M_{\rm s}=2 \times 10^8 {\rm M}_{\odot}$ and 
$R_{\rm s}=0.1$ kpc.
}
\label{Figure. 9}
\end{figure*}

\subsubsection{Mean stellar mass density ($\rho_{\rm s}$)}

As shown in Fig. 8,
the models with higher $\rho_{\rm s}$ ($\rho_{\rm e}$) are likely
to have larger $f_{\rm nrs}$. Furthermore,
there is a threshold $\rho_{\rm s}$ ($\approx 0.1 {\rm M}_{\odot}$ pc$^{-3}$)
above which $f_{\rm nrs}$
can be larger than 0.01: this threshold is referred to as $\rho_{\rm s, th}$
(or $\rho_{\rm e, th}$ at $R_{\rm e}$). Therefore, we can consider that
if stellar systems meet the following condition, then
$f_{\rm nrs} >0.01$  is possible:
\begin{equation}
\rho_{\rm s}  > \rho_{\rm s, th}.
\end{equation}
The models with $M_{\rm s}=10^9 {\rm M}_{\odot}$ 
and $R \ge 3$ kpc can have 
very small $f_{\rm nrs}$  ($<0.002$), which implies that low surface
brightness dwarf galaxies are unlikely to have  NRS. 
For an illustrative purpose, we derive $f_{\rm nrs}$ as a function of $\rho_{\rm s}$
using the results for the fiducial models and those  in which
model parameters other than $R_{\rm s}$ are the same as those
in the fiducial model (shown as big filled circles in Fig. 8).
The functional form of $f_{\rm nrs}$ 
can be written as follows:
\begin{equation}
f_{\rm nrs} = 0.0083 \log \rho_{\rm s} + 0.02.
\end{equation}
This can be rewritten for $\rho_{\rm e}$ as follows:
\begin{equation}
f_{\rm nrs} = 0.0083 \log \rho_{\rm e} + 0.0075.
\end{equation}

Fig. 9 describes the star formation histories of models with
different $\rho_{\rm s}$ for $f_{\rm g}=0.003$,
$t_{\rm a}=1$ Gyr, and $\Delta t_{\rm a}=0.5$ Gyr. 
Clearly, SFRs in lower $\rho_{\rm s}$ are systematically lower
owing to the low mass densities of the newly developed gas disks embedded in
spherical systems.  In some models with rather low $\rho_{\rm s}$,
no star formation is possible (i.e., no NRS). The low-mass models
with ${\rm M}_{\rm s}=3 \times 10^7 {\rm M}_{\odot}$ can show
star formation (but an order of $\approx 10^{-3} {\rm M}_{\odot}$ yr$^{-1}$) 
only if they are very compact ($R_{\rm s} \le 300$ pc).
The UCD model  
with ${\rm M}_{\rm s}=2 \times 10^8 {\rm M}_{\odot}$ 
and $R_{\rm s}=100$ pc shows relative high SFR and larger 
$f_{\rm nrs}$ (0.015). This result suggests that UCDs are likely
to constrain NRS in their central regions.
The model 
with ${\rm M}_{\rm s}=2 \times 10^8 {\rm M}_{\odot}$ 
and $R_{\rm s}=1$ kpc,
however,  shows a much lower SFR and a lower $f_{\rm nrs}$ than the UCD model
which confirms that $\rho_{\rm s}$ is a key parameter for NRS formation.

\subsubsection{Initial gas mass fraction ($f_{\rm g}$)}

Fig. 10 demonstrates that the strongest peaks in [N/Fe] distributions
can be around [N/Fe]=0.5 only in the models with $f_{\rm g} \le 0.01$.
In gas-rich models with $f_{\rm g} > 0.01$,
the vast majority of new stars are formed from original N-poor gas particles
that cannot be so heavily polluted by AGB ejecta. Consequently,
only gas-poor models can have a large fraction ($>0.5$)
 of NRS in their new stars.
The total mass of new stars can be larger than $10^8 {\rm M}_{\odot}$
in the very gas-rich model with $f_{\rm g}=0.1$,
however, 
$R_{\rm nrs}$ is  small ($\approx 0.2$).
The mean SFR is high ($> 0.1 {\rm M}_{\odot}$ yr$^{-1}$)
in this model
so that the adopted assumption of no SNe cannot be realistic
according to the $m_{\rm u}$-SFR relation (W13).
It is likely that ejecta with low [N/Fe] from SNe II can 
significantly lower [N/Fe] of ISM in this gas-rich model.
The model  with $f_{\rm g}$ and $r_{\rm d}=16$ pc show
a much smaller fraction of NRS with [N/Fe]$>1$
than SA1.

\subsubsection{Disk vs spheroid: in situ formation}

Fig. 11 describes the  [N/Fe] distributions of ``in situ'' 
spheroidal and disk models in which
the Galactic bulge is assumed to be formed in a single burst. 
The massive spheroidal stellar systems
with ${\rm M}_{\rm s}= 10^{10} {\rm M}_{\odot}$ 
and $R_{\rm s}=3.2$ kpc can have a large fraction of NRS among all
new stars ($R_{\rm nrs}=0.92$) and their [N/Fe] distributions are 
similar to that of the fiducial model. There is no significant difference
in the [N/Fe] distributions and $f_{\rm nrs}$ ($\approx 0.01$)
between these models with and without
dark matter, which again confirms that dark matter cannot influence
the formation of NRS.
The dependence of [N/Fe] distributions on $t_{\rm a}$, $f_{\rm g}$,
and $\rho_{\rm s}$  in
these models are very similar to those derived for the models
with $M_{\rm s}=10^9 {\rm M}_{\odot}$. 
Intriguingly,
the low-density model with $R_{\rm s}=10$ kpc (corresponding to
a LSB) shows a peak around [N/Fe]$\approx 0.3$ and $f_{\rm nrs}=0$.
All new stars are formed from original gas particle polluted by 
AGB ejecta, though the total mass is only $6 \times 10^5 {\rm M}_{\odot}$.

The disky stellar systems can form NRS within their central regions
for the models with
with ${\rm M}_{\rm s}= 10^{10} {\rm M}_{\odot}$ 
and $R_{\rm s}=3.2$ and finally have
$f_{\rm nrs} \approx 0.01$. This suggests that the formation of NRS is possible,
even if the Galactic bulge was formed from a bar instability in an initially
thin disk. These disky models also show  parameter dependences
of the result
on $t_{\rm a}$ and $f_{\rm g}$ that are similar to those found in
the spheroidal models. 
The low-density model with $M_{\rm s}=10^9 {\rm M}_{\odot}$ (yet
the same mass of dark matter as the above DA1) and $R_{\rm s}=3.2$ kpc
shows a single peak around [N/Fe]=0.8 and a very small total mass
of new stars ($6.2 \times 10^5 {\rm M}_{\odot}$
corresponding to $f_{\rm nrs}=6.2 \times 10^{-4}$). This suggests that
although LSBs can have NRS,  their possibly very low $f_{\rm nrs}$
would make it hard for observers to find them.

\begin{figure*}
\psfig{file=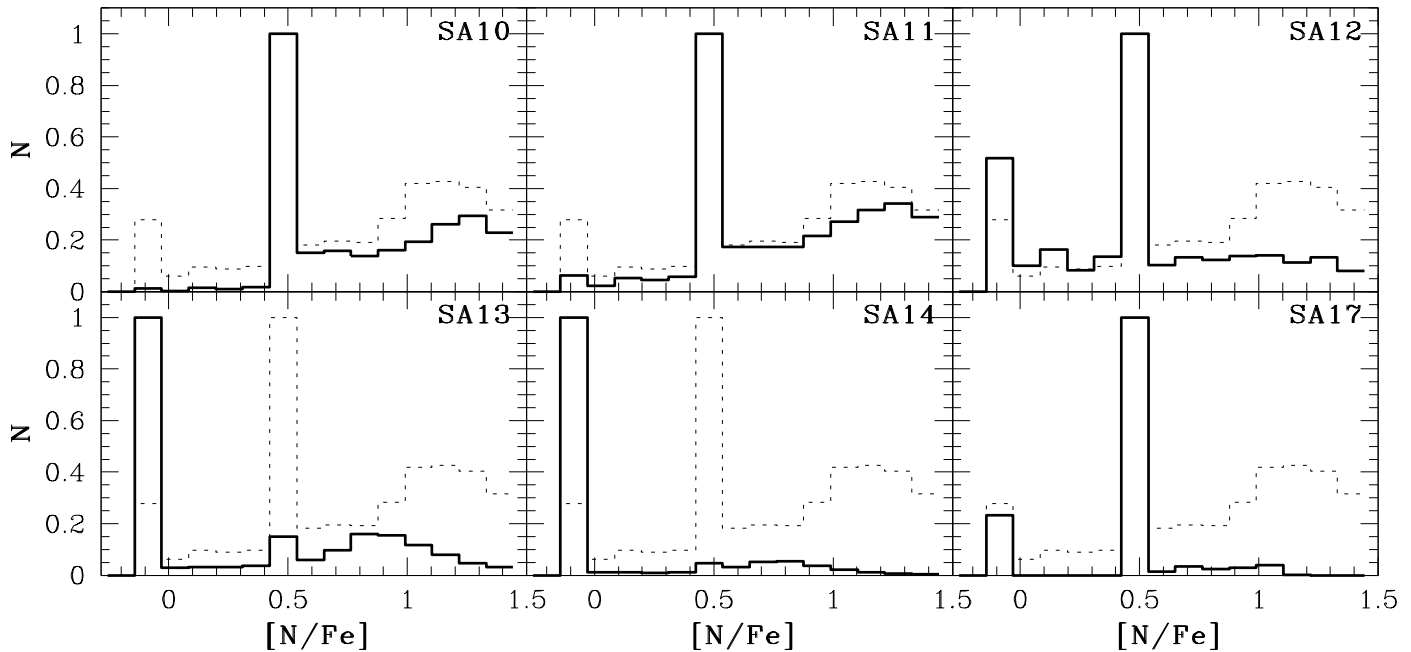,width=18.0cm}
\caption{
The same as Fig. 5 but for models with different $f_{\rm g}$ and 
$r_{\rm d}$.
Model ID is shown in the upper right corner of each panel.
SA10, 11, 12, 13, and 14 are the models with different $f_{\rm g}$
(0.0003, 0.001, 0.01, 0.03, and 0.1, respectively),
and SA17 is the model in which dilution of AGB ejecta by N-poor ISM
is not included ($r_{\rm d}=0$).
In  all models, the adopted star formation histories of the stellar 
systems are identical.
}
\label{Figure. 10}
\end{figure*}

\begin{figure*}
\psfig{file=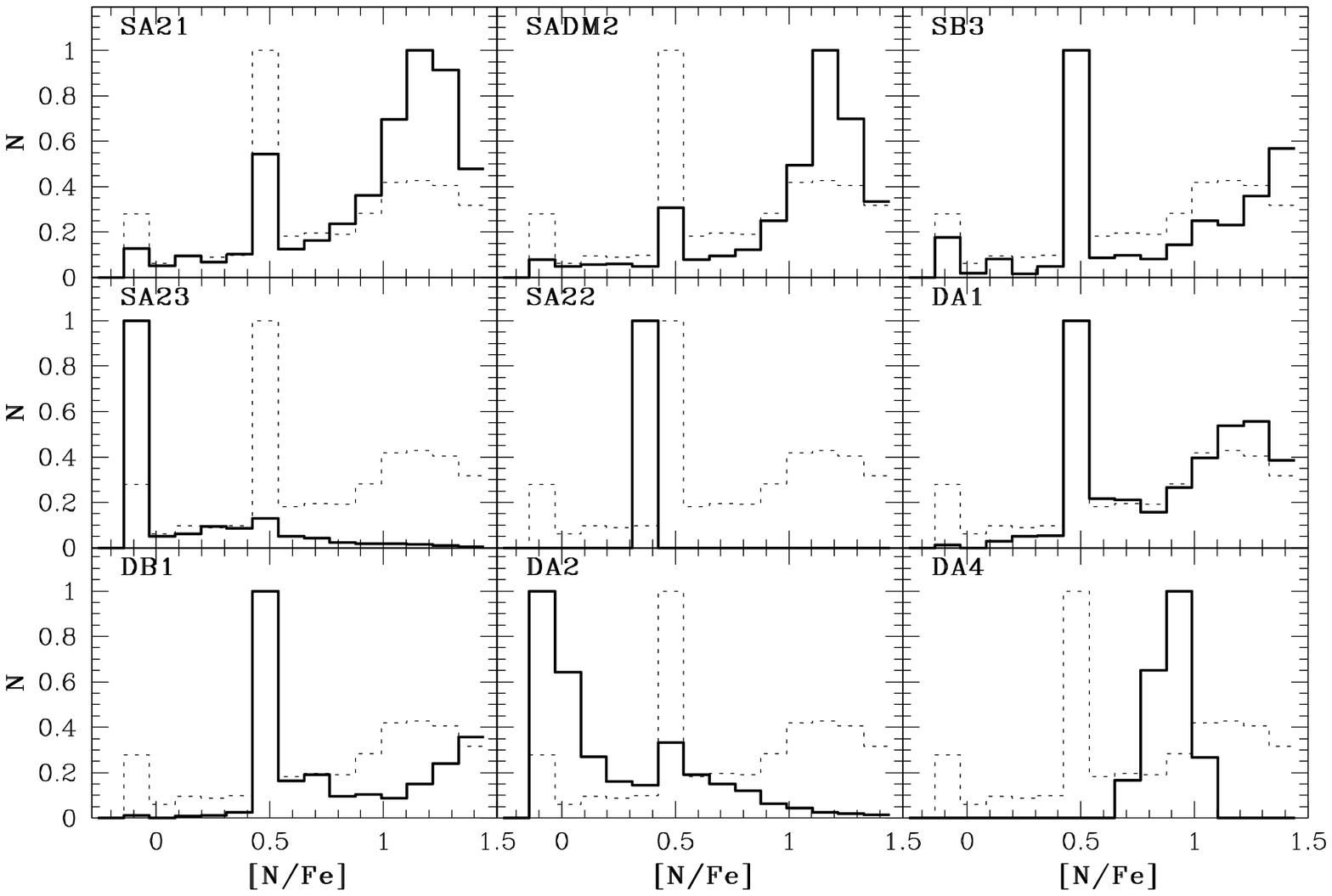,width=18.0cm}
\caption{
The same as Fig. 5 but for ``in situ formation''
models.
SA21, SADM2, SB3, SA23, and SA22 are spheroidal models
with  ${\rm M}_{\rm s}=10^{10} {\rm M}_{\odot}$
(comparable to the present-day bulge mass)
whereas DA1, DB1, DA2, and DA4 are disky models
with ${\rm M}_{\rm s}=10^{9} {\rm M}_{\odot}$
or ${\rm M}_{\rm s}=10^{10} {\rm M}_{\odot}$.
Although SA23 and DA2
with large $f_{\rm g}$ (0.1) and SA22 with low stellar
density ($R_{\rm s}=10$ kpc) have smaller fractions
of NRS, other models show that 
NRS are the dominant populations within the systems.
The spherical models with  (SADM2)  and without 
dark matter (SA21) show intriguing two peaks in their
[N/Fe] distributions.
}
\label{Figure. 11}
\end{figure*}

\section{Discussion}

\subsection{Effects of SNe Ia on [N/Fe] distributions}

Ejecta from SNe Ia can significantly change [Fe/H] of ISM
in the building blocks of the Galactic bulge, if it is gravitationally
trapped by the blocks and then mixed with ISM. However,
the present study did not consider star formation of such mixed gas 
at all. Accordingly, we here discuss how SNe Ia can possibly change the
[N/Fe] of NRS formed from ISM by assuming that (i) stellar systems
have a  canonical IMF (as adopted in the present study),
(2) 3\% of stars with $3 {\rm M}_{\odot} \le m \le 8 {\rm M}_{\odot}$
become binary stars that explode as SNe Ia,
and (ii) $0.8 {\rm M}_{\odot}$ iron is ejected from one SNe Ia.
For an adopted canonical IMF, about $6 \times 10^{-4}$ SNe Ia
is expected for $1 {\rm M}_{\odot}$ stellar system. 
This is lower than $2.7 \times 10^{-3} {\rm M}_{\odot}^{-1}$
estimated for the Large Magellanic Cloud (LMC; Maoz \& Badenes 2010).
If we adopt 
$6 \times 10^{-4} {\rm M}_{\odot}^{-1}$,
the total mass of iron ejected from all possible SNe Ia over the Hubble time
in a stellar system with $M_{\rm s}=10^9 {\rm M}_{\odot}$
is  $4.8 \times 10^5 {\rm M}_{\odot}$ stellar system.
If the total mass of the ISM ($M_{\rm g}$) in the stellar system
is $10^8 {\rm M}_{\odot}$ and if its metallicity is [Fe/H]$=-1$,
then the total mass of iron in the ISM 
is about $1.2 \times 10^4 {\rm M}_{\odot}$.

Therefore, if only  $\approx 2.5$\% ($f_{\rm mix}=0.025$) 
of ejecta from all possible SNe Ia is mixed with
the ISM, then the total iron mass in the ISM
can become by a factor of $\approx 2$ larger than the original mass:
[N/Fe] of stars formed from the ISM can be significantly decreased. 
If $M_{\rm g}$ is $10^7 {\rm M}_{\odot}$, then  $f_{\rm mix}=0.0025$
is required to significantly increase [N/Fe] of stars.
However, it should be stressed here that (i) only a  fraction of all SNe Ia
can explode in less than 1 Gyr corresponding to the formation timescale
of NRS in the present scenario and (ii) SNe Ia ejecta with high
ejection velocity  is unlikely to be 
trapped by gravitational potentials of low-mass stellar building blocks.
We can estimate the fraction of SNe Ia that explode within less than 1 Gyr
($F_{\rm Ia}$) 
by adopting the delay time distributions for SNe Ia ($t_{\rm Ia}$)
adopted in our previous chemical
evolution models of the LMC (Bekki \& Tsujimoto 2012).
If $t_{\rm Ia} \propto (t/{\rm 0.5 Gyr})^{-1}$,
then $F_{\rm Ia} \approx 0.1$. However, 
If the prompt SNe Ia model
is adopted  ($t_{\rm Ia} \propto (t/{\rm 0.1 Gyr})^{-1}$),
then $F_{\rm Ia} \approx 0.5$. Accordingly,
only [10-50]\% of ejecta from all SNe Ia can be mixed with ISM within
$\approx 1$ Gyr.

Since the escape velocities of stellar systems with masses less
than $10^{10} {\rm M}_{\odot}$ that are considered in
the present study  is much lower than 
$\approx 1000$ km s$^{-1}$ (an order of magnitude for the ejection velocity of 
gas from SNe Ia), 
it is possible  that most ejecta from SNe Ia can escape from the systems.
Therefore, it is likely that only a tiny fraction of ejecta from all SNe Ia
can be mixed well with ISM in low-mass stellar systems.
It is not possible for the present study to present a more quantitative
discussion on how much amount of iron ejecta from SNe Ia can be mixed
with ISM without hydrodynamical simulations for the mixing of SNe Ia ejecta
with ISM.
It is thus our future study to numerically
investigate  how [N/Fe] distributions of low-mass stellar
systems can be influenced by SNe Ia.

\subsection{The origin of NRS in the Galactic bulge}

We have demonstrated that $f_{\rm nrs}$ 
in  stellar systems can be as high as $\approx 0.01$,
which is only slightly lower than the observed
for the Galactic bulge  by S17 ($\approx 0.02$),
though there are required conditions for $f_{\rm nrs} \approx 0.01$.
Furthermore, a few models (e.g., SB1 and DB1 with $t_{\rm a}=0.5$ Gyr)  
can show up to $f_{\rm nrs} \approx [0.03-0.04]$.
Accordingly, the present study strongly suggests that NRS in the bulge
can originate from NRS initially in its building blocks: the Galactic 
bulge was formed from massive stellar clumps ($>10^7 {\rm M}_{\odot}$)
initially with NRS.
Since NRS can be formed in the building blocks 
{\it as field stars}, there is no/little need
to invoke a large number of dissolved GCs with NRS in explaining 
the observed $f_{\rm nrs}$.
It is possible that N-rich 2G stars originating from dissolved GCs can be 
a minor population in NRS of the bulge.

However, $f_{\rm nrs}$ can be as high as $\approx 0.01$ 
after $\approx 0.5$Gyr continuous  formation  of NRS in stellar
systems. It could be possible that SNe II or SNe Ia can truncate 
the star formation or reduce [N/Fe] of ISM from where NRS can be formed.
If such truncation of star formation occurs, then
$f_{\rm nrs}$ can be significantly reduced. This possibly low $f_{\rm nrs}$
can be avoided if the IMF of NRS is top-light and bottom-heavy.
For example, if the IMF of a  host stellar system of NRS is a top-heavy
with $\alpha=-1.9$, 
$m_{\rm l}=0.1 {\rm M}_{\odot}$, and
$m_{\rm u}=50 {\rm M}_{\odot}$,
and if that of NRS is a bottom-heavy one with
$\alpha=-2.5$, 
$m_{\rm l}=0.1 {\rm M}_{\odot}$, and
$m_{\rm u}=50 {\rm M}_{\odot}$,
the mass fraction of  low-mass stars with $m \le 0.8 {\rm M}_{\odot}$
among all stars formed
is 0.27 for the host system and 0.68 for NRS.
Accordingly the present-day
$f_{\rm nrs}$ 
can be $\approx 0.01$ even if it was $\approx 0.004$
at the time of NRS formation.
Thus, the possibly low $f_{\rm nrs}$ would not be a serious problem
in the present scenario.

The observed [N/Fe] distribution of the bulge stars shows 
a peak around [Fe/H]$\approx -1$, which is not consistent with
the peak location of the MDF of the Galactic GC system (S17).
Although the present study has not yet provided a clear explanation
for the peak location,
it is possible that this [Fe/H]$\approx -1$ reflects the typical
metallicity of the bulge's building blocks.
It is beyond the scope of this paper to quantify the mean and dispersion
of [Fe/H] among the  stars of  the bulge's building blocks.
Chemical enrichment can proceed in high-density ISM,
because  more efficient star formation in denser gas
can  end up with more rapid
chemical enrichment.
Therefore,
if only high-density building blocks can form NRS in the present
scenario, the typical metallicity of the building blocks could be 
higher.

If star formation from ISM polluted heavily by AGB stars can continue
even after stars with $m \le 3 {\rm M}_{\odot}$ start to pollute
ISM, then ISM within stellar systems can have higher 
abundances of $s$-process elements due to the high abundance in the AGB
winds
(e.g., [La/Fe]$\approx 2$ for $m=2 {\rm M}_{\odot}$ in  F14).
Accordingly, if such low-mass AGB stars contribute  to the chemical
pollution of ISM, then NRS can possibly show significant enhancement
in [La/Fe] and [Ba/Fe]. If only more massive AGB stars can contribute
to the chemical enrichment, then the level of such $s$-process enhancement
should be minor (e.g., [La/Fe]$\approx 0.6$ for $m=5 {\rm M}_{\odot}$; F14).
Currently, it is not observationally clear whether NRS have higher
[La/Fe] and [Ba/Fe] in comparison with N-normal/N-poor bulge stars.
Thus,  [La/Fe] and [Ba/Fe] in NRS combine to provide
strong constraints on (i) a mass range of AGB stars that
contribute to the chemical enrichment of ISM by AGB winds
and therefore (ii) the formation timescale  of NRS.

In the present study, we have focused exclusively on the AGB scenario,
but did not discuss the OB wind scenario at all. As shown in BC07,
the fraction of N-rich gas ([N/Fe]$>0.5$) can be high ($\approx 0.5$),
however, such N-rich gas cannot be converted efficiently into new stars within
star-forming molecular clouds owing to the strong feedback effects of
SNe II.  If this N-rich gas is recycled into ISM (without being expelled
by SNe II) in massive and dense stellar systems with much
deeper gravitational potentials than molecular clouds,
then NRS could be formed
from such gas. This possible retention of N-rich gas originating from
star-forming clouds would needs to be investigated in our future studies.
NRS formed from such  N-rich gas cannot show abundance
enhancement in $s$-process elements as shown in the AGB scenario:
[La/Fe] and [Ba/Fe] of NRS can be used to distinguish between 
the two scenarios for NRS formation.

\subsection{Formation of NRS in the Galactic stellar halo}

The Galactic stellar halo is observed to have N-rich 
(``CN-strong'') stars with
the fraction being $\approx 0.026$ among all stars of the halo
(e.g., MG10; Koch et al. 2019). The similarity
in chemical abundances between the NRS and 2G stars in the Galactic GCs
has led astronomers to think that their origin is related to GC destruction
in the Galaxy. However, as suggested by the above authors,
a large number of GCs with 2G stars should be destroyed to explain the
observed significant fraction of NRS. If the total mass of the halo
is $10^9 {\rm M}_{\odot}$ and if the mass fraction of 2G stars
in GCs is typically 0.5 for a typical GC mass
of $2 \times 10^5 {\rm M}_{\odot}$, 
then $f_{\rm nrs}=0.026$ means that $\approx 260$
GCs should be {\it completely} destroyed. 
Since the required number of GCs is
significantly larger than the currently observed one ($\approx 150$),
this GC destruction scenario cannot properly explain the observed
$f_{\rm nrs}$. Furthermore, NRS are observed to exist even in the outer part
of the halo ($R>30$ kpc), where tidal destruction of GCs is highly unlikely.
Thus, this scenario is not so promising for the origin of NRS in the halo.

The present new results suggest that if most of the halo's building blocks
have mean stellar densities higher than the required 
threshold stellar density ($\rho_{\rm s, th}$) for
the formation of NRS, then the observed $f_{\rm nrs}$ can be naturally
explained in the present scenario. 
We accordingly suggest that the vast majority of NRS in the halo can 
originate not from GCs but
from its high-density building blocks initially with NRS.
The presence of NRS in the outer halo region ($R>30$ kpc)  can be due to tidal
stripping of stars of the high-density building blocks.
Intriguingly, $f_{\rm nrs}$ is very
similar between the bulge and the halo,
which can be also naturally explained by the present scenario.

\subsection{Stellar densities of galaxies as a key parameter for NRS}

The present study has shown that $f_{\rm nrs}$ can be 
significant ($>0.01$) only in
stellar systems with $\rho_{\rm s} > \rho_{\rm s, th} 
\approx 0.1 {\rm M}_{\odot}$ pc$^{-3}$, which means
$\rho_{\rm e}$ should be higher than
$\rho_{\rm  e, th} \approx 3.2 {\rm M}_{\odot}$ pc$^{-3}$
at the half-mass radii.
This threshold stellar density ($\rho_{\rm e, th}$) can be
used to discuss whether dwarf spheroidal  galaxies (dSph) 
can form NRS based on the observed size-mass relation
(e.g., Fig. 1 in Forbes et al. 2014).
Most of dSphs have 100 pc$<R_{\rm e}<$1000 pc
and 
$10^5 {\rm M}_{\odot} < M_{\rm s}< 10^7 {\rm M}_{\odot}$,
which means $1.2 \times 10^{-5}  {\rm M}_{\odot}$ pc$^{-3}$ 
 $<\rho_{\rm e}<$
$1.2 {\rm M}_{\odot}$ pc$^{-3}$.
Although only massive ($M_{\rm s} > 10^7 {\rm M}_{\odot}$ 
and more compact ($R_{\rm e} < 2$ kpc) dSphs 
can have $\rho_{\rm e}$ higher than $\rho_{\rm e, th}$,
such dSphs are rare (e.g., F14): only two dSphs appear to have
$\rho_{\rm e} > \rho_{\rm e, th}$.
Therefore, it is unlikely that dSphs can have $f_{\rm nrs} \approx 0.01$.

Although Recent observational studies indeed have tried to find NRS in nearby
dwarf galaxies  (e.g., Sculptor dwarf spheroidal),
they did not find NRS (e.g., Salgado et al. 2019).
Hasselquist et al. (2017) have found that only one of many stars in Sagittarius
dwarf galaxy
shows [N/Fe]$\approx 0.6$ (see Fig. 3 in their paper).
These results are consistent with the prediction of the present study,
because the stellar mass densities of these dwarf spheroidal galaxies
are rather low.
Compact elliptical galaxies (cEs) and UCDs are highly likely to have
$f_{\rm nrs}>0.01$, because they have quite small $R_{\rm e}$ 
(F14). Massive ($M>10^9 {\rm M}_{\odot}$)
yet compact ($R<1$ kpc) dwarf elliptical galaxies (dEs)
are also likely to have $f_{\rm nrs}>0.01$,
and indeed there are a number of such objects seen in F14.
Although spectroscopic confirmation of the presence of NRS in these
galaxies with high $\rho_{\rm e}$  is formidable,   
such observational studies are valuable for the origin of NRS.

A remaining question is whether stellar systems with any masses
can have a significant fraction of NRS as long as they have
$\rho_{\rm s} > \rho_{\rm s, th}$.
If ${\rm M}_{\rm s}$ is lower than $10^6 {\rm M}_{\odot}$,
then, AGB winds cannot be retained is stellar systems so that
new stars cannot be formed from AGB ejecta (B11).
On the other hands, if stellar systems are massive enough to retain
a large amount of
warm/hot gas ($T>10^6$ K) within them,
then AGB ejecta will be thermally heated up due to interaction with
such warm/hot gas. Accordingly, it would be unlikely that the AGB ejecta
can finally form local high-density regions from where NRS can be formed.
Accordingly, there could be a mass range for which NRS can be formed from
AGB ejecta after its mixing with ISM.
Fig. 12 briefly illustrates which stellar systems are more likely 
to have significant fractions of NRS ($f_{\rm nrs}>0.01$).

\subsection{NRS in  elliptical galaxies}

Giant elliptical galaxies with $r$-band
magnitude ($M_{\rm r}$) of $\approx -22$ mag
are observed to have [N/Fe]$\approx 0.2$
and [N/Fe] is observed to be higher for smaller $M_{\rm r}$
(i.e., brighter elliptical galaxies; Schiavon 2007).
These [N/Fe] values  for elliptical galaxies
are based on the 
integrated spectroscopic results for the entire galaxies.
Therefore, even if mass fractions
of  NRS are significant ($\approx 0.01$)  in these galaxies,
the average [N/Fe] can be small, because the vast majority of the
stars have lower [N/Fe]. If an elliptical galaxy consists of NRS with
[N/Fe]=1 and N-normal stars with [N/Fe]=0 and if their mass
fractions are 0.05 and 0.95, respectively,
the average [N/Fe] is $\approx 0.15$. Therefore, the observed
[N/Fe]=0.2 implies that bright elliptical galaxies can have
a large fraction of NRS.
The higher [N/Fe] in more luminous elliptical galaxies 
implies that more luminous elliptical galaxies were formed 
from a larger fraction of high-density galactic building blocks
initially with NRS.

Recent theoretical and observational studies have suggested that
the ultraviolet (UV) upturn of giant elliptical galaxies can be closely
related to disintegration of 
massive GCs that are building blocks of the galaxies,
because such GCs can have He-rich stars.
(Bekki 2012; Goudfrooij 2018).
Since  NRS should have also high $Y$
in the present scenario, 
they can be also responsible for the origin of the UV upturn that
can be caused by a significant fraction of He-rich stars.
If giant elliptical galaxies were formed from multiple merging of high-density
galactic building blocks, they should be able to have a significant fraction
of He-rich stars too. 
Therefore, the origin of the UV-upturn is closely related to the
formation of NRS within the building blocks of giant elliptical galaxies.

\begin{figure}
\psfig{file=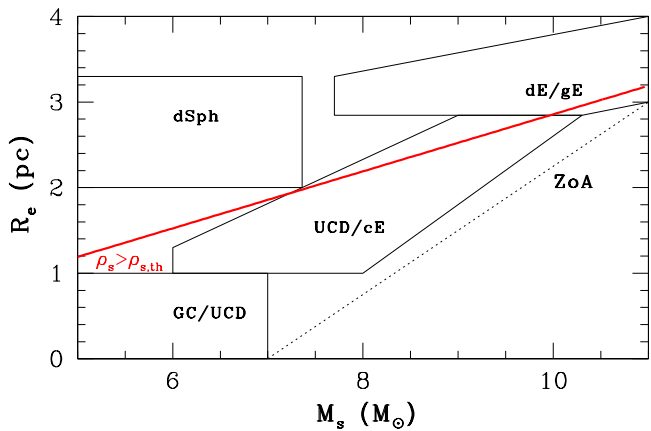,width=8.5cm}
\caption{
A schematic figure that describes which stellar systems can possibly
have NRS with $f_{\rm nrs} >0.01$. Stellar systems that are located below
the red solid line, where $\rho_{\rm s}=\rho_{\rm s, th}$, can have
$f_{\rm nrs} > 0.01$. The Zone of Avoidance (ZoA)
where there are no galaxies
is below the black dotted line.
The observed locations of different type of galaxies
(GC, UCD, dSph, cE, dE, and gE) are based on the observational results 
by Forbes et al. (2014).
It should be stressed here that $\rho_{\rm s, th}$ is estimated for
the building blocks of galaxies, not for the present-day stellar systems
that are shown in this figure.
}
\label{Figure. 12}
\end{figure}

\section{Conclusions}

In order to elucidate the origin of N-rich stars with [N/Fe]$>0.5$ (NRS) observed
in the Galactic bulge, we have adopted a new scenario of NRS
formation and thereby investigated in what physical conditions such
stars can be formed. In the scenario,  NRS were first formed
in the building blocks of the bulge  (e.g., stellar clumps; Noguchi 1998;
Elmegreen et al. 2013) and then
become the members of the bulge after merging of these building blocks.
We have first investigated whether NRS can be formed from 
interstellar medium (ISM) mixed
with N-rich ejecta from (i) stellar winds of massive OB stars
or (ii)  asymptotic giant branch (AGB) stars and found that AGB ejecta is more promising in
explaining the observed fraction of NRS in the bulge.

Therefore, using our original numerical simulations with a new mixing
model of AGB ejecta,
we have  investigated the details of  star formation processes
from ISM polluted by 
AGB stars in massive
spherical and disky stellar systems that are the building blocks of 
the bulge. Since we have investigated this for the first time,
we have assumed that (i) a  stellar system is formed
in a single starburst 
and (ii) new stars can be formed after most of the original gas
is swept away by energetic Type II supernovae (SNe II) events during
the formation of the systems.
In this preliminary study,
we  have adopted an assumption that there are no SNe Ia and SNe II events during
the formation of NRS.
Non-inclusion of these SNe events would lead the present study to
over-estimate the fraction of NRS ($f_{\rm nrs}$), because iron production
from these SNe can decrease [N/Fe] of stars formed from ISM polluted by
such events.

The key parameters in this first parameter study are
(i) the total masses of 
the stellar systems ($M_{\rm s}$),  (ii) the  mean stellar densities 
($\rho_{\rm s} \approx M_{\rm s}/R_{\rm s}^3$, where $R_{\rm s}$ is
the size of a galaxy), (iii) the gas mass fraction of ISM ($f_{\rm g}$),
(iv) the mean ages of the stellar populations ($t_{\rm a}$), 
and (v) the duration of the initial starbursts ($\Delta t_{\rm a}$). 
We have mainly investigated how the mass fractions of NRS
among all stars ($f_{\rm nrs}$) depend on these key parameters and how
the simulated [N/Fe] can be determined by  the parameters.
The principal results are as follows:

(1) NRS can be formed in spherical stellar systems 
with masses of $10^8-10^9 {\rm M}_{\odot}$ when 
an enough amount of AGB ejecta is accumulated within the systems
so that the ISM densities can be rather high ($[10^3-10^5]$ atom cm$^{-3}$).
This formation process of NRS from ISM heavily polluted by AGB ejecta
is essentially the same as that of NRS within  GCs
(e.g., B07; D'Ercole et al. 2010).
However, the required $\rho_{\rm s}$ 
($0.1 {\rm M}_{\odot}$ pc$^{-3}$)
for NRS formation
is much lower than
the typical GC density ($\rho_{\rm s}=2.4 \times 10 {\rm M}_{\odot}$ pc$^{-3}$
for 
$M_{\rm s}=2 \times 10^5 {\rm M}_{\odot}$ and $R_{\rm s}=10$ pc
or $R_{\rm e}=2$ pc).
Accordingly, GCs are not the only formation sites of NRS.
The systems should have smaller $f_{\rm g}$ ($<0.03$) so that the new
stars can be dominated by NRS.
The [N/Fe] distributions in the new stars  is quite diverse depending on
the mixing processes of AGB ejecta with N-poor ISM.

(2) The total masses of NRS formed in spherical stellar systems
over $\approx 0.5$ Gyr timescale of star formation 
can be larger than $\approx 1$\% of the total stellar masses, 
if their $\rho_{\rm s}$ are  
larger than $\approx 0.1 {\rm M}_{\odot}$ pc$^{-3}$.
(or $\rho_{\rm e} > 3.2 {\rm M}_{\odot}$ pc$^{-3}$).
This means that the mass fractions of NRS among all stellar populations
can be significant
only in high-density galactic building
blocks, and accordingly
they are unlikely to be  formed in low-mass dwarf spheroidal and elliptical
galaxies on their size-mass relation.
If the stellar initial mass function (IMF) during the formation of NRS
is top-light or steeper than that for their host stellar systems,
then the present-day mass fraction of low-mass NRS
($m \le 0.8 {\rm M}_{\odot}$) can be as high as or even
higher than $\approx 0.03$. \\

(3) Star formation rates (SFRs) during the formation of NRS
are rather low ($\approx [10^{-3}-10^{-2}] {\rm M}_{\odot}$ yr$^{-1}$) 
in models with low $f_{\rm g}$ $<0.01$.
Theoretical predictions on the relation between SFRs and the maximum
possible mass of stars ($m_{\rm u}$)
in  star formation (W13) imply that
the IMF during the formation
of NRS can be rather top-light ($m_{\rm u}<25 {\rm M}_{\odot}$; a smaller
number of massive SNe II).
Collisions of molecular clouds that can trigger massive star formation
(e.g., Fukui et al. 2017) are also unlikely to occur in gas-poor systems.
It is therefore unlikely that N-poor ejecta ([N/Fe]$<0$) from SNe II
can significantly lower [N/Fe] of ISM from where NRS can be formed, even
if ISM is polluted by such ejecta. \\

(4) Although NRS can be formed in gas-rich stellar systems
($f_{\rm g} > 0.1$), the mass fraction of NRS among new
stars is rather small ($<0.1$) due to efficient dilution of N-rich ejecta
by a large amount of N-poor ISM.  
SFRs in such gas-rich systems are rather high, and cloud-cloud
collisions that trigger massive star formation is highly likely within the systems.
Accordingly
$m_{\rm u}$ can be also high, which ends up with a larger mass
fraction of SNe II. This  implies that
ejecta with low [N/Fe] from
SNe II can significantly lower [N/Fe] of ISM in gas-rich dwarfs
with high SFRs.
Therefore, if such chemical enrichment by SNe II is included in
the present gas-rich models, then the formation of NRS should be 
severely suppressed. It is thus highly unlikely that gas-rich dwarf galaxies
can have NRS. \\

(5) NRS can be formed within  disks with $M_{\rm s}=10^{10}$
embedded in massive dark matter
halos (``in situ formation''), 
and $f_{\rm nrs}$ can be higher than $0.01$ in some models
with  low $f_{\rm g}$.  
These suggest that if the Galactic bulge was formed from an initially
thin disk through bar instability, it can have a fraction of NRS
in its inner region ($R<500$ pc).
Both spherical and disky stellar systems show rather compact distributions
and rotational kinematics in NRS. 
It is yet to be understood how the kinematics of NRS looks like,
if the bulge was formed from merging of such high-density stellar systems
with central NRS. Recent observational results on the kinematics
of NRS (e.g., Fern\'andez-Trincado et al. 2019a, b;
Savino \& Posti 2019) will be able to be used to
discuss this point in our future study.
\\

(6) The present results for disky and spherical systems thus suggest
that NRS in the bulge can be formed either from their 
high-density building blocks
(e.g., stellar clumps) with
$M_{\rm s}= [10^8-10^9] {\rm M}_{\odot}$
or from a stellar disk: there is no need to
invoke the destruction of too many GCs ($>200$) 
with the so-called ``second generation'' (2G) stars with high [N/Fe]
in explaining the origin of [N/Fe].
We suggest that although 2G stars in GCs can become NRS in the bulge,
the contribution of such GC destruction processes in the formation
of NRS of the bulge
can be minor. \\

(7) NRS observed in the Galactic halo can be formed from 
the destruction of high-density
building blocks (dwarf-like galaxies)
initially with NRS. Given that the mass fraction
of NRS can be as large as 0.02,
the observed mass/number fraction of NRS in the halo can be readily explained by
the formation of the halo by merging of the building blocks.
Accordingly,  the destruction of 
a larger number ($>200$) of GCs with
2G stars with high [N/Fe] is not required to explain 
the origin of NRS in  the halo. \\

(8) The formation processes of NRS observed
in ultra-compact dwarfs (UCDs)
and giant elliptical galaxies can be essentially the same as those
for the Galactic bulge. One of the prediction  from the present model
is that NRS in the bulge, UCDs, and elliptical galaxies should
have higher $Y$. The observed He-rich stars in the Galactic bulge
(e.g., Nataf \& Gould 2012) and the UV-upturn 
phenomena in elliptical galaxies, which can be caused helium enhancement
(e.g., Ali et al. 2018; Goudfrooij 2018),  can have the same origin as NRS.\\

(9) The present study did not investigate $f_{\rm nrs}$ of stellar systems
in the context of the OB wind scenario, because $f_{\rm nrs}$ is likely to be
very low owing to the removal of N-rich ejecta from ISM by SNe II. 
However, unlike GC-forming molecular clouds,
massive building blocks of the bulge
can possibly retain  N-rich ISM polluted by OB winds 
under some physical conditions
(e.g., deeper  gravitational potentials). Accordingly,
the OB wind scenario would also need to be investigated for the 
better understanding of the origin of NRS. \\

\section{Acknowledgment}
I (Kenji Bekki; KB) am   grateful to the referee  for  constructive and
useful comments that improved this paper.

\end{document}